\documentclass[aps,prd,amsmath,nofootinbib,twocolumn,10pt,showpacs,preprintnumbers]{revtex4-1}

\usepackage{mystyle}
\usepackage{pgfkeys}

\newcommand{\nrmaxphaseerror}{0.2}
\newcommand{\nrmaxamplitudeerror}{5}
\newcommand{\nrmaxunfaith}{7\times 10^{-4}}
\newcommand{\tcircval}{30}
\newcommand{\maxFreqAmpDiffPercent}{4}
\newcommand{\imrconstructioncase}{16}
\newcommand{\minnrmassearly}{80}
\newcommand{\minnrmassdesign}{230}
\newcommand{\xmatch}{0.11}
\newcommand{\xblend}{0.12}
\newcommand{\numnewsims}{20}
\newcommand{\numoldsims}{3}
\newcommand{\numsims}{23}

\newcommand{\typicalcyclesrefpeak}{7}

\newcommand{\optional}[1]{}

\newcommand{\code}[1]{{\tt #1}}
\newcommand{\aligoOOne}{\textit{aLIGO O1}}
\newcommand{\aligoDesign}{\textit{aLIGO design}}

\newcommand{\circpercent}{\maxFreqAmpDiffPercent\%}
\newcommand{\xref}{\xmatch}

\newcommand{\tcirc}{t_{\mathrm{circ}}}

\newcommand{\tpeak}{t_{\mathrm{peak}}}
\newcommand{\trel}{t_{\mathrm{rel}}}

\newcommand{\eccparamsnot}{(x_0,e_0,l_0,\phi_0)}

\newcommand{\MSun}{M_{\odot}}
\newcommand{\tref}{t_{\text{ref}}}

\newcommand{\Hz}{\text{ Hz}}
\newcommand{\xblendsym}{x_{\text{blend}}}
\newcommand{\xrefsym}{x_{\text{ref}}}
\newcommand{\re}{\text{Re}}

\pgfkeys{/cases/39/number/.initial=1}
\pgfkeys{/cases/15/number/.initial=2}
\pgfkeys{/cases/54/number/.initial=3}
\pgfkeys{/cases/21/number/.initial=4}
\pgfkeys{/cases/22/number/.initial=5}
\pgfkeys{/cases/16/number/.initial=6}
\pgfkeys{/cases/18/number/.initial=7}
\pgfkeys{/cases/20/number/.initial=8}
\pgfkeys{/cases/19/number/.initial=9}
\pgfkeys{/cases/17/number/.initial=10}
\pgfkeys{/cases/45/number/.initial=11}
\pgfkeys{/cases/62/number/.initial=12}
\pgfkeys{/cases/61/number/.initial=13}
\pgfkeys{/cases/43/number/.initial=14}
\pgfkeys{/cases/41/number/.initial=15}
\pgfkeys{/cases/59/number/.initial=16}
\pgfkeys{/cases/44/number/.initial=17}
\pgfkeys{/cases/40/number/.initial=18}
\pgfkeys{/cases/53/number/.initial=19}
\pgfkeys{/cases/50/number/.initial=20}
\pgfkeys{/cases/49/number/.initial=21}
\pgfkeys{/cases/48/number/.initial=22}
\pgfkeys{/cases/52/number/.initial=23}

\newcommand{\caselabel}[1]{\begingroup
\pgfkeysgetvalue{/cases/#1/number}{\temp}%
{Case \#\temp}%
\endgroup%
}

\newcommand{\doubleline}{\hline \hline}

\usepackage{printlen}

\begin{document}

\title{An eccentric binary black hole inspiral-merger-ringdown
  gravitational waveform model from numerical relativity and
  post-Newtonian theory}

\newcommand{\CIFAR}{\affiliation{Canadian Institute for Advanced Research, Toronto M5G 1Z8, Canada}} %
\newcommand{\CITA}{\affiliation{Canadian Institute for Theoretical Astrophysics, University of Toronto, Toronto M5S 3H8, Canada}}
\newcommand{\Cornell}{\affiliation{Cornell Center for Astrophysics and Planetary Science, Cornell University, Ithaca, New York, 14853, USA}}

\newcommand{\AEI}{\affiliation{
Max-Planck-Institut f\"ur Gravitationsphysik,
Albert-Einstein-Institut, \\ 
Am M\"uhlenberg 1, D-14476 Golm, Germany}
}

\author{Ian Hinder}
\AEI

\email{ian.hinder@aei.mpg.de}

\author{Lawrence E. Kidder}
\Cornell

\author{Harald P. Pfeiffer}
\CITA \AEI \CIFAR

\date{\today}

\newcommand{\omfitrefrelres}{1}

\newcommand{\typicalcycles}{20}
\newcommand{\maxinitecc}{0.2}

\newcommand{\maxecc}{0.08}
\newcommand{\maxq}{3}

\newcommand{\targetfaith}{97}
\newcommand{\targetunfaith}{3}

\newcommand{\maxmtarget}{85}       

\newcommand{\lowmass}{70}          
\newcommand{\lowmassecc}{0.05}     
\newcommand{\lowmassq}{3}          

\begin{abstract}
We present a prescription for computing gravitational waveforms for
the inspiral, merger and ringdown of non-spinning eccentric binary black hole systems.  
The inspiral waveform is computed using the post-Newtonian
expansion and the merger waveform is computed by interpolating a
small number of quasi-circular NR waveforms.
The use of circular merger waveforms
is possible because
eccentric binaries circularize in the last few cycles before the
merger, which we demonstrate up to mass ratio $q = m_1/m_2 = 3$.  
The complete
model is calibrated to $\numsims{}$ numerical relativity (NR)
simulations starting $\approx \typicalcycles{}$ cycles
before the merger with eccentricities $e_{\text{ref}} \le \maxecc{}$ and
mass ratios $q\le3$, where $e_{\text{ref}}$ is the eccentricity $\approx \typicalcyclesrefpeak{}$
cycles before the merger.  The NR waveforms are long enough that they start above
$30 \Hz$ ($10 \Hz$) for BBH systems with total mass $M \ge \minnrmassearly \MSun$ ($\minnrmassdesign
\MSun$).
We find that, for the sensitivity of advanced LIGO at the time of its first observing run,
the eccentric model has a
faithfulness with NR of over $\targetfaith{}\%$ for systems with total
mass $M\ge\maxmtarget{}M_{\odot}$ across the parameter space
($e_{\text{ref}}\le\maxecc{}$, $q\le\maxq{}$).
For systems with total mass $M\ge\lowmass{}M_{\odot}$, the
faithfulness is over $\targetfaith{}\%$ for $e_{\text{ref}}\lesssim\lowmassecc{}$
and $q\le\lowmassq{}$.
The NR waveforms and the Mathematica code for the model are publicly
available.
\end{abstract}

\maketitle{}



\section{Introduction}

In 2015, LIGO detected the gravitational wave (GW) event GW150914
corresponding to the merger of a binary black hole (BBH)
system~\cite{Abbott:2016blz}.  Subsequently, further events from BBH mergers
have been detected~\cite{Abbott:2016nmj,Abbott:2017vtc}. The parameters of
the binaries were inferred
from the measured data using waveform models calibrated to numerical relativity
(NR) simulations~\cite{Taracchini:2013rva,Khan:2015jqa,Mroue:2013xna}
under the very reasonable assumption that the orbit
of the binary was quasi-circular~\cite{TheLIGOScientific:2016wfe}.  This is expected
because binary eccentricity
decays quickly under the emission of gravitational
radiation~\cite{Peters:1964zz}.

Whilst there is no known mechanism by which a BBH system could retain
a non-negligible eccentricity in the last $\sim 4$ orbits before
merger that LIGO was able to see for the high mass event GW150914,
we would like to confirm this
astrophysical prediction by comparing the data to general relativistic
waveforms including eccentricity.  Further, several scenarios have
been suggested in which binaries may retain non-negligible
eccentricity for an extended
time~\cite{2003ApJ...598..419W,Benacquista:2002kf,Gultekin:2004pm,1976ApJ...204L...1T,Blaes:2002cs,Dotti:2005kq,Antonini:2015zsa,Rodriguez:2016kxx,Maccarone:2007dd,Strader:2012wj,Chomiuk:2013qya,Rodriguez:2015oxa},
including some where the binary may enter the sensitive frequency band
of LIGO or LISA a large number of cycles before merger, and before this
eccentricity has decayed.

Gravitational wave data analysis using the method of matched filtering
requires accurate models of the waveforms in order to measure source
parameters.  For the inspiral of eccentric binaries, models based on
the post-Newtonian (PN) approximation can be used.  PN theory for binaries
in eccentric orbits is very well
developed~\cite{PhysRev.131.435,Peters:1964zz,Wagoner:1976am,Blanchet:1989cu,1992MNRAS.254..146J,Blanchet:1993ec,Rieth:1997mk,Damour:1988mr,Schaefer93,1995CQGra..12..983W,Damour:2004bz,2006PhRvD..73l4012K,Memmesheimer:2004cv,Arun:2007sg,Arun:2007rg,arunblanchetam},
but because the approximation assumes that the black holes are widely-separated
and slowly-moving, PN can only model the inspiral waveform.

Near the merger,
the waveform can only be determined using full numerical solutions of
the Einstein equations.  Numerical relativity results for eccentric
BBH systems were first presented in
\cite{Sperhake:2007gu,Hinder:2007qu}.  In \cite{Hinder:2007qu}, it was
found that the eccentricity in equal-mass eccentric binary waveforms becomes irrelevant
a few cycles
before the merger. At this point, to a good approximation, the binary 
circularizes leaving the same merger waveform and a black hole of the
same final mass and spin as if the inspiral had been circular.

The first comparison between eccentric PN waveforms and NR was
performed in \cite{Hinder:2008kv}, and agreement was found
between 21 and 11 GW cycles before
merger.  Comparisons at later times were compromised by
inaccuracy in the NR simulations.
In \cite{Mroue:2010re} and \cite{Lousto:2015uwa}, the evolution of
eccentricity in NR simulations was compared with Newtonian
and PN predictions and shown to be broadly consistent.

While it is possible to perform NR simulations of BBH mergers for a
small number of configurations, each simulation takes several weeks to
run, and is too computationally expensive to use for GW parameter
estimation, where typically millions of waveforms with different
parameters are generated and compared to the data.  Further, the
length of NR simulations is generally limited to tens of orbits, also
due to computational expense.  Therefore, computationally inexpensive waveform
models are required, which reproduce the NR waveforms to a sufficient accuracy, and
which can produce waveforms with the large numbers of orbits which may
be visible to a gravitational wave detector.

The first eccentric waveform model incorporating
inspiral, merger and ringdown (IMR) was presented in
\cite{Huerta:2016rwp}.  The inspiral is based on PN, using an
improved version of the ``$x$-model''
\cite{Hinder:2008kv}, and the merger is modeled by the
non-eccentric \textit{Implicit Rotating Source} model of \cite{Kelly:2011bp}
assuming that eccentricity will be
negligible by the time of the merger.
The complete IMR model is called the ``$ax$-model''.  An initial comparison with two
NR waveforms showed that the model was realistic.
Recently, two different models for eccentric binary
inspiral-merger-ringdown waveforms for nonspinning BBH systems in the
Effective-One-Body framework were proposed.  In
\cite{Hinderer:2017jcs}, the foundations of an eccentric model were
presented.  In \cite{Cao:2017ndf}, another model was developed, and
initial comparisons were performed with three NR simulations.

In this paper, we present a new set of \numnewsims{} eccentric non-spinning NR
simulations using the Spectral Einstein Code (SpEC), which we have
made available as part of the public catalog \cite{sxscatalogue} of the Simulating
eXtreme Spacetimes collaboration.  The simulations have initial
eccentricities $e \le \maxinitecc{}$, mass ratios
$q\le3$ and generally start $\approx \typicalcycles{}$ cycles before
the merger.  The new simulations show that the circularization shortly
before merger observed in \cite{Hinder:2007qu} for equal-mass binaries
extends to binaries with mass ratio $q\le3$.  This justifies the use
of circular merger waveforms in \cite{Huerta:2016rwp}.

Independently of \cite{Huerta:2016rwp}, we develop an
eccentric IMR model based on the PN $x$-model~\cite{Hinder:2008kv}, combined with a
circular merger waveform.  Our quasi-circular merger waveform, which can be evaluated
for any mass ratio $q$ within the calibration region $1 \le q \le 4$,
is obtained by interpolation between several NR waveforms with
different $q$.  For the transition region between the
eccentric PN portion and the circular NR portion, we use a
prescription calibrated to the new eccentric NR simulations, for which
an essential ingredient is a fit of the time between the waveform
reaching a given reference frequency and the peak of the
waveform amplitude.

We test our model against the NR simulations, quantifying the
agreement in phase, amplitude, and faithfulness (a measure of how
close the waveforms are when observed by a gravitational wave
detector).  Since the waveforms do not agree perfectly, there is an
ambiguity in the choice of PN parameters to use when comparing with a
given NR waveform.  We choose PN parameters such that the waveforms
agree shortly before the merger, and measure the accumulation of error
at preceding times.  This allows us to be confident in the behavior
of the model at the merger, and we expect that improvements to the
PN inspiral model would extend its validity to earlier times.

The NR waveforms are available in the SXS Public Waveform Catalog
\cite{sxscatalogue}, and a Mathematica package \code{EccentricIMR}
implementing the full inspiral-merger-ringdown model is available as
open source software at \cite{eccentricimrcode}.


In Sec.~\ref{sec:pnmodel}, we identify the main features of the
eccentric PN $x$-model which is the basis of our IMR model.
In Sec.~\ref{sec:nrsims}, we describe the NR simulations, including
the code used and the eccentric configurations that we simulated.  We
validate the waveforms by assessing the main sources of error, and
finally we show that the circularization observed in
\cite{Hinder:2007qu} extends to mass ratios $q\le3$.
Sec.~\ref{sec:measuringecc} discusses the method we use to define
eccentricity in NR. 
Sec.~\ref{sec:circmodel} shows how the circular merger model
(CMM) is constructed by interpolating between NR waveforms.
In Sec.~\ref{sec:constructimr}, we explain how the IMR model is
constructed by combining the PN inspiral with the CMM, and the
calibration to NR simulations, including the time-to-merger fit, that
we use for the transition.
In Sec.~\ref{sec:fouriercompare}, we address issues related to
computing Fourier transforms of eccentric NR waveforms, required for
computing the faithfulness of waveforms for gravitational wave data
analysis.
In Sec.~\ref{sec:modellingresults}, we compare the IMR
model with the NR waveforms, analyzing both the time-domain
frequencies, phases and amplitudes, as well as the Fourier-domain
faithfulness relevant to gravitational wave detection and parameter
estimation.
Finally, Sec.~\ref{sec:conclusions} summarizes our results, and
discusses possible future improvements of our model.

Throughout, we use units in which $G = c = 1$.

\section{PN inspiral model}
\label{sec:pnmodel}

In this section, we briefly review the ``$x$-model'', an eccentric PN
inspiral model introduced in \cite{Hinder:2008kv} and involving a
change of variables of the ``$n$-model'' presented in
\cite{2006PhRvD..73l4012K}.  We begin by recalling Newtonian eccentric
orbits.

Consider the relative orbit of two bodies at positions $\vec x_1$,
$\vec x_2$ of masses $m_1$ and $m_2$.  We restrict to the case where
the bodies orbit in the $xy$ plane, since in the non-spinning case,
BBH systems will orbit in a plane due to symmetry.  The separation $r
= |\vec x_1 - \vec x_2|$ satisfies
\begin{eqnarray}
r &=& a \left [ 1-e \cos u \right ] \label{eq:rofu}
\end{eqnarray}
where $a$ is the semi-major axis of the orbit, $e$ is the
eccentricity, which parametrizes the amplitude of the oscillations in
$r$, and $u$ is the {\em eccentric anomaly}, an angular variable which
represents the phase of the oscillation in $r$.  Pericenter (point of closest approach)
is at $u=0$, and $u=\pi$ corresponds
to apocenter.  The angular velocity of the orbit is given by
\begin{eqnarray}
\dot \phi &=& \frac{n \sqrt{1-e^2}}{\left [ 1-e \cos u \right ] ^2} \label{eq:phidotofu}
\end{eqnarray}
where $n$ is the {\em mean motion}, defined as $2\pi/P$, where $P$,
the radial period, is the time between pericenter passages.  In
Newtonian dynamics, the quantities $a$, $e$, $n$ and $P$ are
constants, which can all be expressed in terms of the energy $E$ and
angular momentum $L$; i.e.~only two of them are independent.  
$u$ can be determined by solving the Kepler equation,
\begin{eqnarray}
l &\equiv& 2\pi(t-t_0)/P = u - e \sin u \, ,\label{eq:kepler}
\end{eqnarray}
a transcendental algebraic equation for $u$, where $l$ is called the
{\em mean anomaly}, and $t_0$ is a time corresponding to pericenter
passage. $l$ parametrizes the time elapsed since the preceding
pericenter passage.  Eq.~\ref{eq:kepler}
can be solved numerically for $u$, for example by Newton's method, at
each $t$.  Thus we can obtain $r$ and $\dot \phi$ (and hence $\dot r$
and $\phi$) at any time $t$.  Each orbit can be parametrized by the
four independent constants $a$, $e$, $\phi(\bar t)$ and $l(\bar t)$ at
a time $t = \bar t$.

The post-Newtonian \textit{quasi-Keplerian representation} of
eccentric orbits~
\cite{Wagoner:1976am,Blanchet:1989cu,1992MNRAS.254..146J,Blanchet:1993ec,Rieth:1997mk,Damour:1988mr,Schaefer93,1995CQGra..12..983W,Gopakumar:1997bs,Gopakumar:2001dy,Memmesheimer:2004cv}
is based on the above description, and utilizes the
same quantities.  However, the equations relating these quantities contain post-Newtonian
corrections, expressed as powers of $v/c$.  For a pedagogical introduction
to the post-Newtonian Kepler problem, see \cite{poisson2014gravity}.  
In the PN description, there are three different eccentricities,
$e_t$, $e_r$ and $e_{\phi}$, related to each other by PN
expressions. These are introduced to simplify certain
relations. Here, we express everything in terms of $e \equiv e_t$.
Relativistic eccentric
orbits have the new feature of precession of the pericenter.  The azimuth of the pericenter
increases by $\Delta \phi$ during one radial (pericenter to pericenter) period $P$, so
that the azimuthal coordinate $\phi$ increases by
$2\pi + \Delta \phi$ during this time.
$n$ no
longer reduces to the angular velocity $\dot \phi$ in the circular
limit, and instead we introduce the quantity
$\omega\equiv(2\pi+\Delta\phi)/P$, the average angular velocity.
In the relativistic \textit{circular} case,
$\omega = \dot \phi$.  For Newtonian orbits, further, $\omega = n$.
When radiation reaction effects are neglected, $\omega$ is a constant.
The choice of PN variables (for example $n$ or $\omega$) to use to
expand the equations is arbitrary, and leads to different \textit{approximants} once the PN series is truncated.  In \cite{Hinder:2008kv}, it was
found that expanding the equations in $x = (M \omega)^{2/3}$ led to
better agreement between NR and PN than expanding them in $n$, as had
been customary previously.  The $\omega$ variable has the benefit of agreeing
with the angular velocity used as an expansion variable in
the quasi-circular Taylor T4 model which has been shown to agree well with NR in the non-spinning equal-mass case~\cite{Boyle2007}, but
we know of no deep reason why $\omega$ (equivalently
$x$) should be better than $n$.  We expect this good agreement to
deteriorate for spinning systems or for $q >> 1$~\cite{MacDonald:2012mp}.
We parametrize the orbit in
terms of the two dimensionless quantities $x$ and $e$.  

In the PN model, Eq.~\ref{eq:rofu} remains unchanged, but
Eqs.~\ref{eq:phidotofu}--\ref{eq:kepler} are modified by PN correction
terms, expanded to 3 PN order.  Relativistic
orbits not only differ by pericenter precession and PN correction
terms, but the energy and angular momentum, which in the Newtonian
case are constants, also change due to the emission of gravitational
waves.  In the \textit{adiabatic approximation}, the fluxes of $E$ and
$L$ are approximated by time-averaging over a radial period $P$, and
are used to calculate the time derivatives of $x$ and $e$.  These fluxes are
used here to 2 PN order.

Hence, in order to compute an orbit subject to energy and angular
momentum loss, it is first necessary to solve the pair of coupled ODEs
for $\dot x$ and $\dot e$, then compute $u$ using the PN Kepler
equation, from which $l$, $r$ and $\phi$ are obtained.

Finally, we compute the gravitational wave using the
\textit{restricted approximation}, in which the $\ell=2,m=2$
spin-weight $-2$ spherical harmonic mode of the waveform is given to
leading (quadrupolar, Newtonian) order as 
\begin{align}
h^{{2}{2}} &= \int {}_{-2} {Y^2_2}^*(\theta,\varphi) h(\theta,\varphi) d\Omega \\
&= - \frac{4 M \eta e^{ -2i \phi}}{R} \sqrt{\frac{\pi}{5}} \left ( \frac{M}{r} + (\dot \phi r + i \dot r)^2 \right ) \, , \label{eqn:h22} \\
h^{{2}{-2}} &= h^{22*} \, .
\end{align}
Here, ${}_{-2} Y^2_2(\theta,\varphi) = \frac{1}{2} e^{2 i \varphi
}\sqrt{5/\pi} \cos ^4\left(\theta/2\right)$, and $\theta$ and $\varphi$
are the spherical polar angles of
the observer.  The $\ell=2,m=\pm 2$ modes dominate for small $e$. 

For the purpose of this work, we consider the $x$-model as a black
box, completely defined in \cite{Hinder:2008kv}, that produces
$h^{22}(t)$ for a given $(x_0,e_0,l_0,\phi_0)$.  The model is expected
to be a good approximation of the relativistic dynamics when the
separation is large and the velocity is small, and will break down
close to the merger.

\section{NR simulations}
\label{sec:nrsims}

\subsection{NR methods}

\code{SpEC}~\cite{SpECwebsite,Scheel:2006gg,Szilagyi:2009qz,Buchman:2012dw}
is a pseudo-spectral code capable of efficiently solving many types of
elliptic and hyperbolic differential equations, with the primary goal
of modeling compact-object binaries.  For smooth problems, spectral
methods are exponentially convergent and high accuracy can be achieved
even for long simulations.  \code{SpEC} evolves the first order
formulation~\cite{Lindblom:2005qh} of the generalized harmonic
formulation of Einstein's
equations~\cite{Friedrich1985,Pretorius2005a}. The damped harmonic
gauge~\cite{Lindblom:2009tu} is used to provide stable coordinate
conditions. Singularities inside BHs are dynamically excised
from the computational domain using feedback control systems
~\cite{Hemberger:2012jz,Scheel:2014ina} and initial conditions of
low orbital eccentricity are obtained by an iterative evolution
procedure~\cite{Pfeiffer:2007yz}. \code{SpEC} uses h-p adaptivity\footnote{h-p adaptivity
refers to varying both the size, $h$, of the elements and the order, $p$, of
the polynomials in each element.}
to dynamically control numerical truncation error and to increase
computational efficiency~\cite{Szilagyi:2014fna}. Waveforms are
extracted using the Reggie-Wheeler-Zerilli formalism
on a series of coordinate spherical shells and extrapolated to null
infinity using polynomial expansions in powers of the areal
radius~\cite{Boyle:2009vi}.

\subsection{Configurations}

We aim to simulate BBH configurations with a given initial frequency parameter
$x$, eccentricity $e$ and mean anomaly $l$ (see Sec.~\ref{sec:pnmodel}).
We use the PN approximation to translate these quantities into the initial data
parameters needed by \code{SpEC}, namely the orbital angular velocity
$\dot\phi$, the separation of the horizon centroids $r$, and the
radial velocity $\dot r$.
This specification of initial data parameters is only an approximation, because the
PN and NR quantities are expressed in different coordinate systems,
and the NR initial data contains non-astrophysical {\em junk
  radiation} which perturbs the parameters of the binary away from
those given in the initial data.  Nevertheless, we find that this
prescription gives waveforms which agree to a good approximation with
the PN $(x,e,l)$.  These initial data parameters are not
used at any point in the subsequent analysis; all quantities are measured from the
waveforms, so any discrepancy is not important.

\begin{table}
\begin{tabular}{llllllll}
  \doubleline
  Case & Simulation & $q$ & $x_0$ & $e_\text{ref}$ & $l_\text{ref}$ & $t_\mathrm{peak}$ & $N_\mathrm{orbs.}$\\
\hline
1 & SXS:BBH:0180 & 1 & 0.0540 & 0.00 & 0.667 & 8720.2 & 26.7 \\
2 & SXS:BBH:1355 & 1 & 0.0718 & 0.05 & -2.801 & 2551.6 & 11.9 \\
3 & SXS:BBH:1356 & 1 & 0.0582 & 0.07 & 0.931 & 6001.0 & 20.8 \\
4 & SXS:BBH:1357 & 1 & 0.0689 & 0.10 & 1.344 & 2889.0 & 12.8 \\
5 & SXS:BBH:1358 & 1 & 0.0703 & 0.10 & -1.789 & 2655.9 & 12.1 \\
6 & SXS:BBH:1359 & 1 & 0.0711 & 0.10 & 2.727 & 2530.5 & 11.7 \\
7 & SXS:BBH:1360 & 1 & 0.0710 & 0.14 & 2.093 & 2372.7 & 11.1 \\
8 & SXS:BBH:1361 & 1 & 0.0713 & 0.14 & 1.469 & 2325.5 & 10.9 \\
9 & SXS:BBH:1362 & 1 & 0.0710 & 0.19 & 0.905 & 2147.2 & 10.2 \\
10 & SXS:BBH:1363 & 1 & 0.0711 & 0.19 & 0.500 & 2108.8 & 10.1 \\
11 & SXS:BBH:0184 & 2 & 0.0710 & 0.00 & -0.604 & 3014.6 & 13.7 \\
12 & SXS:BBH:1364 & 2 & 0.0697 & 0.05 & 2.132 & 3200.3 & 14.2 \\
13 & SXS:BBH:1365 & 2 & 0.0696 & 0.06 & 1.926 & 3180.8 & 14.1 \\
14 & SXS:BBH:1366 & 2 & 0.0696 & 0.10 & 0.963 & 3073.3 & 13.6 \\
15 & SXS:BBH:1367 & 2 & 0.0702 & 0.10 & -0.743 & 2955.3 & 13.3 \\
16 & SXS:BBH:1368 & 2 & 0.0709 & 0.10 & -2.010 & 2850.1 & 13.0 \\
17 & SXS:BBH:1369 & 2 & 0.0693 & 0.19 & -1.575 & 2616.7 & 11.9 \\
18 & SXS:BBH:1370 & 2 & 0.0710 & 0.19 & 1.691 & 2376.9 & 11.1 \\
19 & SXS:BBH:0183 & 3 & 0.0745 & 0.00 & 1.818 & 2811.9 & 13.5 \\
20 & SXS:BBH:1371 & 3 & 0.0696 & 0.06 & -2.301 & 3707.5 & 16.2 \\
21 & SXS:BBH:1372 & 3 & 0.0696 & 0.09 & 2.963 & 3564.6 & 15.6 \\
22 & SXS:BBH:1373 & 3 & 0.0701 & 0.09 & 1.640 & 3451.5 & 15.3 \\
23 & SXS:BBH:1374 & 3 & 0.0695 & 0.18 & -0.481 & 3014.9 & 13.5 \\

  \doubleline
\end{tabular}
\caption{Eccentric NR simulations used in this work.  The columns give
  the case number, the SXS catalog number, the mass ratio $q = m_1/m_2$, where $m_1$
  and $m_2$ are the masses of the black holes, the initial average
  orbital frequency parameter $x_0$, the eccentricity $e_{\text{ref}}$
  and mean anomaly $l_{\text{ref}}$ measured at a reference frequency
  $x_{\text{ref}} = 0.075$, the time since the start of
  the usable waveform at which $|h_{22}|$ reaches its peak, and the
  number of orbits.}
\label{tab:nrsims}
\end{table}

We perform new simulations for \numnewsims{} eccentric non-spinning
configurations using the \code{SpEC} code.  We also use
\numoldsims{} existing quasi-circular non-spinning configurations
already available in the SXS Public Waveform
Catalog~\cite{Mroue:2013xna}.  The parameters of these configurations
are given in Table \ref{tab:nrsims}.  In order to assess the numerical
truncation error, each configuration is run at multiple resolutions.
The error analysis is presented in Sec.~\ref{sec:nrerrors}.  For each of the
cases 1--23, Table
\ref{tab:nrsims} gives the SXS catalog identification number, the mass ratio
$q=m_1/m_2$, the orbital frequency parameter $x_0$ measured after the junk radiation
portion of the waveform,
the 
eccentricity $e_{\text{ref}}$ and mean anomaly $l_{\text{ref}}$ measured at a
reference frequency $x_{\text{ref}} = 0.075$, the time of the peak of the
amplitude of the dominant mode of the gravitational wave strain
$|h_{22}|$, and the number of orbits simulated.  $x$, $e$ and $l$ are measured
entirely from the waveforms by fitting to PN as described in Sec.~\ref{sec:measuringecc}.
$x_{\text{ref}}$ was chosen as the lowest frequency common to all the waveforms, and
for most simulations, corresponds to a time close to the start of the simulation.

\begin{figure}[h]
  \includegraphics{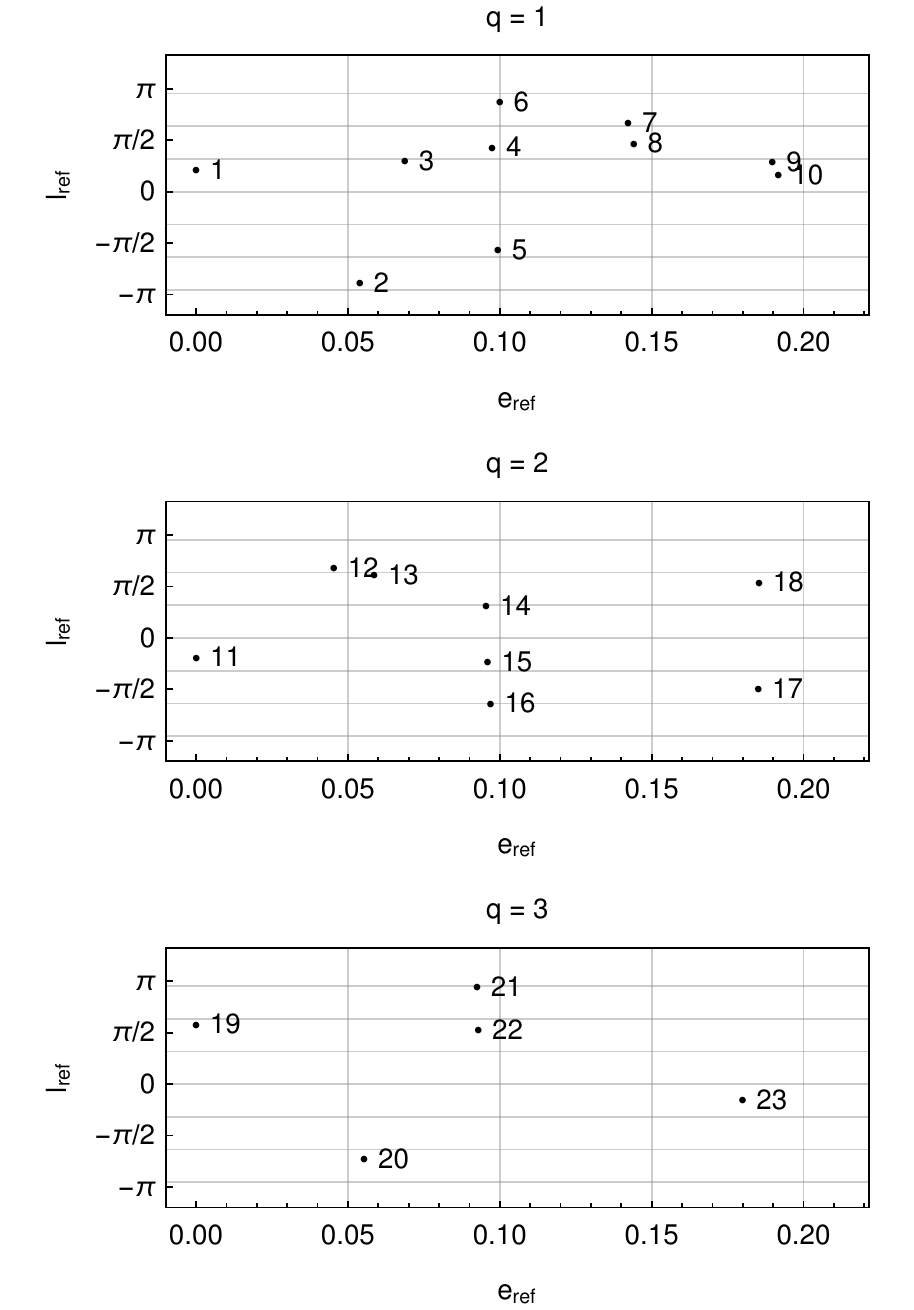}
  \caption{The NR configurations plotted as a function of eccentricity
    $e_{\text{ref}}$ and mean anomaly $l_{\text{ref}}$ at the reference
    frequency $x_{\text{ref}} = 0.075$ for different mass ratios $q$.}
  \label{fig:paramspace}
\end{figure}

Fig.~\ref{fig:paramspace} shows
the distribution of eccentricities and mean anomalies in the parameter space.  The configurations
span mass ratios $q \le 3$, eccentricities $0 \le e_0 \le 0.2$, and mean anomalies $-\pi < l_0 \le \pi$.  Most of the eccentric configurations start
at an average orbital frequency parameter of 
$x \sim 0.07$ 
and evolve
for between 11 and 15 orbits before merging.   

\subsection{Effects of eccentricity in waveforms}

\begin{figure}[h]
  \includegraphics{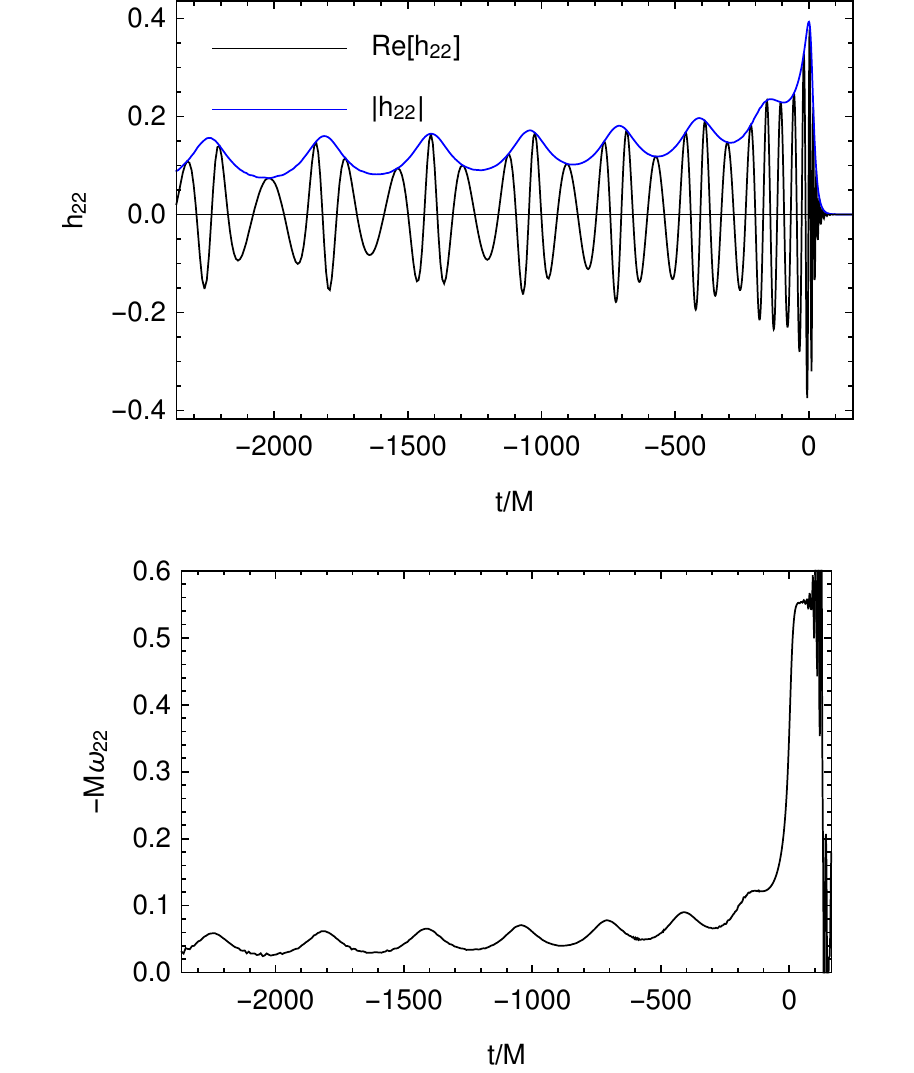}
  \caption{An NR waveform with eccentricity $e_{\text{ref}} = 0.19$ (Case \#9).  The
    top panel shows the real part and amplitude of the dominant
    $\ell=2, m=2$ spherical harmonic mode of the strain, and the
    bottom panel shows the frequency of this mode, computed as
    $\omega_{22}=\frac{d}{dt}{\arg h_{22}}$.  The oscillations in
    $|h_{22}|$ and $\omega_{22}$ are characteristic features of
    eccentricity.}
  \label{fig:waveform_example}
\end{figure}

Fig.~\ref{fig:waveform_example} shows an example of one of the
eccentric waveforms, \caselabel{19}, with $e_{\text{ref}} = 0.19$.
The usual oscillations in the
strain (top panel) at twice the orbital frequency are modulated by an
oscillating envelope with a frequency lower than the orbital frequency, corresponding to
precession of the pericenter.  These modulations due to eccentricity persist at least up
to $\sim 3$ cycles before the merger.  The instantaneous gravitational
wave frequency (bottom panel) also shows oscillations due to
eccentricity, where in the quasi-circular case, the frequency would
vary monotonically.  The period of the oscillations in the amplitude
and instantaneous frequency corresponds to the radial orbital period
$P$, and the amplitude of the oscillations is related to the eccentricity $e$.
The phase of the oscillations is associated with the \textit{mean anomaly}, $l$.
See Sec.~\ref{sec:pnmodel} for the definitions of these quantities.

\subsection{Accuracy of the NR waveforms}
\label{sec:nrerrors}

We have verified that the NR waveforms are not dominated by numerical
truncation error due to finite resolution of the simulations.  The
waveform phase error accumulated up to the peak of $|h_{22}|$, estimated
as the difference between the highest two
resolutions~\cite{Hinder:2013oqa}, is less than $\nrmaxphaseerror$ radians.  The amplitude
error during the inspiral, computed as a function of phase, is
typically below $1\%$, but in a small number of cases is as large
as $\nrmaxamplitudeerror\%$.  The unfaithfulness (see
Sec.~\ref{sec:faithfulness}) between the NR waveforms at different
resolutions is $1-F < \nrmaxunfaith$ in all cases.  Some
configurations, for example those for higher mass ratios, were run
with higher resolution in order to attain this accuracy.

\subsection{Circularization}
\label{sec:circularisation}

\begin{figure*}
  \includegraphics{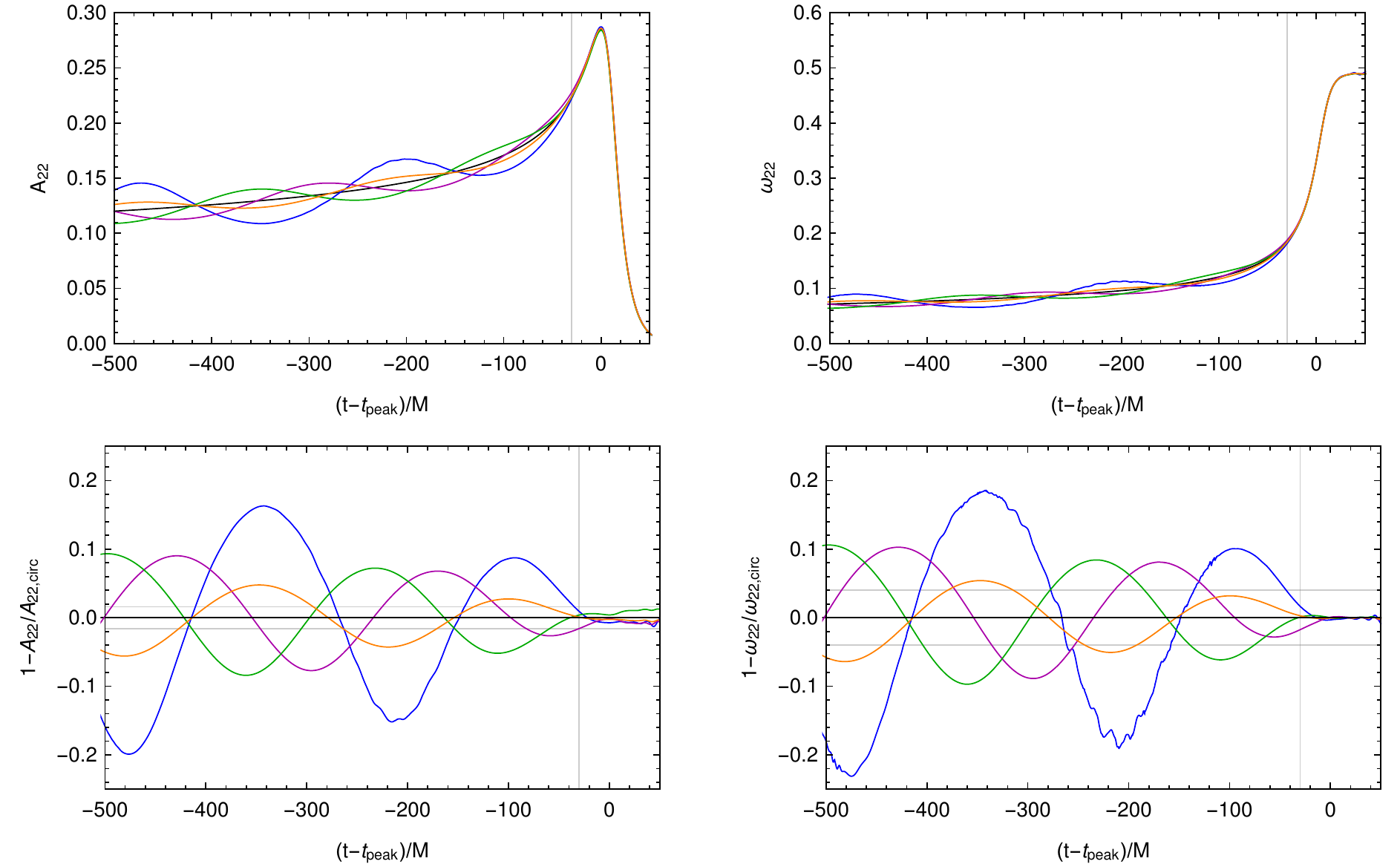}\\
  \caption{Circularization of q=3 non-spinning binary black hole
    waveforms.  Shown in the upper panels are the amplitude, $A$, and
    frequency, $\omega_{\mathrm{22}}$, of the $l=2, m=\pm 2$ mode of the gravitational wave
    strain.  The lower panels show the fractional deviations from the
    non-eccentric results $A_{\mathrm{22,circ}}$ and $\omega_{\mathrm{22,circ}}$.
    The gray horizontal lines in the lower panels
    indicate $\pm \circpercent$.}
  \label{fig:circ_amp_freq}
\end{figure*}

In \cite{Hinder:2007qu}, it was shown that when non-spinning
equal-mass black holes merge, the mass and spin of the resulting black hole were independent
(within numerical error) of the eccentricity of the binary for initial
eccentricities $e \le 0.4$ (measured when the radial period $P \sim 387 M$).  Further, the gravitational wave frequency
from binaries with different initial eccentricities was shown to be
visually indistinguishable for $t > t_{\text{peak}} - 50 M$ (Fig.~3 of
\cite{Hinder:2007qu}).  This provides evidence that the 
binary circularizes within $\sim 50 M$ of the merger.

Here, we study this circularization for \textit{unequal} mass systems
with mass ratios up to $q=3$.  Fig.~\ref{fig:circ_amp_freq} shows the amplitude and frequency of the gravitational wave
for several eccentricities with mass ratio $q = 3$.  We
clearly see the effect of eccentricity as oscillations in both the frequency and
amplitude of the waveform for $t < t_{\text{peak}} - \tcircval M$.
However, for $t > t_{\text{peak}} - \tcircval M$, the waveforms are
visually indistinguishable.  The lower panels of Fig.~\ref{fig:circ_amp_freq} show the relative difference
between each eccentric configuration and the circular case.  We see that
for $t>t_{\text{peak}}-\tcircval{} M$, the waveform amplitude and
frequency differ by only $\circpercent{}$ for all the different
eccentricities at fixed mass ratio $q$.
For $q=1$ and $q=2$, we find the same behavior; irrespective of
initial eccentricity (up to $e_0 \lesssim 0.2$) and initial mean
anomaly, all simulations at the same mass ratio $q$ show nearly
identical amplitude and frequency for $t>t_{\text{peak}}-\tcircval{} M$.

We interpret this to mean that the binary has circularized by
$\tcircval{} M$ before the merger, to an accuracy of $\circpercent$.
Hence, when modeling the waveform from eccentric binaries, there may
be no need to use an eccentric model for the merger portion, as using
a circular model instead may introduce a negligible effect on
observables such as the waveform.  Fig.~\ref{fig:circ_amp_freq}
justifies the use
of circular merger waveforms in \cite{Huerta:2016rwp}.

\section{Measuring eccentricity}
\label{sec:measuringecc}

Given an eccentric waveform model, we may wish to use it to measure
the parameters of a GW signal.   This amounts to determining
$(M,q,x(\tref),e(\tref),l(\tref),\phi(\tref); \tref)$ at some
reference time $\tref$.  Since there is a freedom to choose $\tref$,
we quote the parameters at a fixed value of $x_{\text{ref}}$.  For
example, in Table \ref{tab:nrsims}, we use $x_{\text{ref}} = 0.075$.

In simplified terms, the parameters of a GW source are measured by
comparing the measured GW strain data to the model, and determining
the model parameters which best reproduce the data.  Note that the
measured parameters are therefore PN parameters.  There are plausible
quasi-local GR definitions for black hole masses and spins, and we can
therefore ask what \textit{bias} is introduced in these measured
parameters by using an approximate PN-based model instead of a true GR
(or NR) waveform.  We can determine this by fitting the PN model to an
NR waveform of known masses and spins, and measuring the difference in
the measured parameters.

However, in the eccentric case, the situation is complicated by the
fact that there is no clear general relativistic definition of
eccentricity with which to label an NR waveform (see
\cite{Mroue:2010re} and \cite{Healy:2017zqj} for various possible
definitions).
Our approach is to
define $e$ and $l$ of the NR system as the PN values for which the
agreement between the instantaneous NR and PN waveform frequency,
$\omega_{22} = d/dt \arg(h_{22})$, is maximized over a single radial
period centered on a reference time at which $x = x_{\text{ref}}$.  This
is possible because we find that the PN model we are using, with 3 PN
conservative dynamics, agrees very well with NR over one radial
period, as shown in \cite{Hinder:2008kv}.  The dominant error in our
model is the 2 PN adiabatic evolution of $x$ and $e$ on timescales longer than one
radial period.  If the agreement over one radial period were not good,
then it would be problematic to use PN to define the eccentricity of
an NR waveform.

We fit the PN model to the NR data as follows.  First, we choose a
time window $[t_1,t_2]$ in which to fit the eccentric PN model.  We
then perform a least squares fit of $\omega_{22}^{\mathrm{PN}}(x,e,l)$
to $\omega_{22}^{\mathrm{NR}}$ to determine $(x,e,l)$.  We then perform
an additional fit of $\phi_{22}^{\mathrm{PN}}(x,e,l,\phi)$ to
$\phi_{22}^{\mathrm{NR}}$ to determine $\phi$.  This is the same
procedure used in \cite{Hinder:2008kv}.

This fitting can be performed over any time interval, and gives the
best-fitting PN parameters over that one interval.  Since the NR and
PN waveforms are not the same, the measured parameters and the
resulting waveform will depend on the choice of fitting interval.

In Sec.~\ref{sec:modellingresults}, we will compare the eccentric IMR
model to the NR waveforms.  For this purpose, we choose to fit the PN
model to the NR waveform at $x = \xref$, which typically occurs $\approx \typicalcyclesrefpeak{}$
cycles before the merger.  The time interval used for fitting is
centered on this point with total width equal to the radial period
$P$.  Note that the choice of fitting window therefore depends on $x$
and $P$ from the fit.  We use an iterative process, starting from an
initial guess for the fitting window location and width, and update
the guess based on the result of the fit.
We
use the parameters measured at this point to label the waveform.  The
RMS fit error in each case is $\le 1\%$, indicating that the PN model
accurately describes the waveform on a timescale of one radial period
close to the merger.

\begin{figure}
  \includegraphics{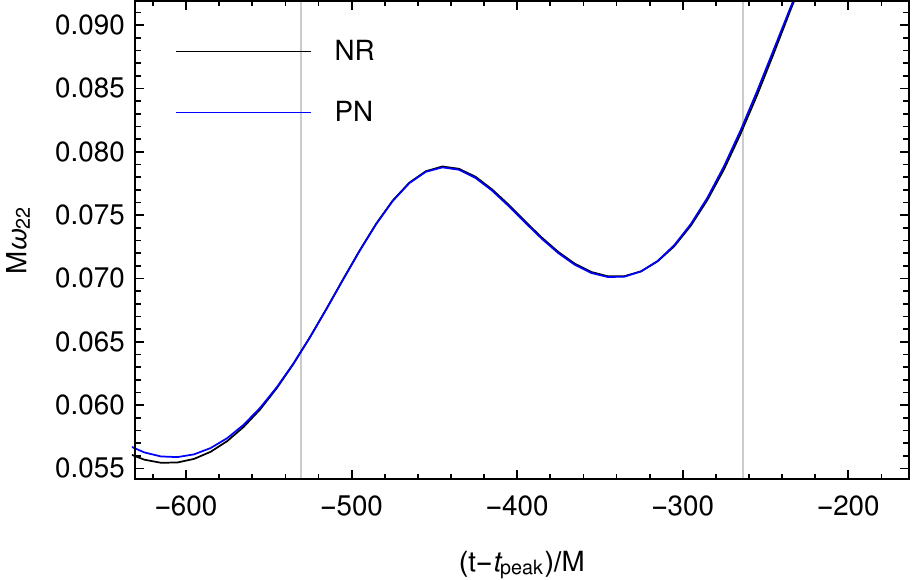}
  \caption{PN fit to NR frequency to measure eccentric parameters at
    $x = \xref$ shortly before merger.}
  \label{fig:nrpncompareom}
\end{figure}

For the configuration with $q = 1$ and $e_0=0.1$ as determined by the
initial parameters (\caselabel{\imrconstructioncase}), the comparison between $\omega^{\mathrm{NR}}$ and
$\omega^{\mathrm{PN}}$ for the $\ell=2,m=2$ mode, fitted across one radial period
at $x=\xref$, is shown in Fig.~\ref{fig:nrpncompareom}.  Note that
this procedure for choosing the PN parameters corresponding to an NR
waveform is not unique.  For example, in \cite{Hinder:2008kv}, a
longer fitting interval near the start of the NR waveform was chosen.
Our motivation in this work is to accurately model the merger, so we
choose to make the NR and PN waveforms agree close to the merger, and
then evaluate the growth in error at earlier times.

\section{Circular merger model}
\label{sec:circmodel}

As shown in Sec.~\ref{sec:circularisation}, the eccentric NR waveforms
circularize before the merger, suggesting that it should be possible
for an eccentric waveform model to incorporate a circular model for
the merger.
Any circular waveform model should be sufficient.  For
example the Implicit-Rotating-Source (IRS) model~\cite{Kelly:2011bp} used in \cite{Huerta:2016rwp},
effective-one-body models such as SEOBNRv4~\cite{Bohe:2016gbl}, or
surrogate models~\cite{Blackman:2015pia} formed by interpolating NR
waveforms.  As shown in Fig.~4 of \cite{Huerta:2016rwp}, the IRS model
does not match the NR data perfectly, and the EOB and surrogate models introduce
additional complications to our model which are not necessary for modeling the straightforward
waveform from a non-spinning BBH merger.  Hence, we created a very simple model for
the merger waveform by performing an interpolation in $q$ of a small
number of non-eccentric non-spinning NR waveforms in the neighborhood
of the merger.  The resulting circular merger model (CMM) can be evaluated
for $1 \le q \le 4$ and arbitrary $\phi_0$, corresponding to the
initial phase of the waveform.  This model agrees well with
non-eccentric NR waveforms.  Note that there is no attempt to ensure validity for $q > 4$,
and it will very likely break down for these mass ratios, though extension of the method
to higher mass ratios should present no difficulties.

To construct the CMM, we take three non-spinning \textit{input}
waveforms from the SXS waveform catalog~\cite{Mroue:2013xna} with
mass ratios 1, 2 and 4 (SXS:BBH:0180, SXS:BBH:0184 and SXS:BBH:0182, respectively),
and apply a time shift such that the peak of
$|h_{22}|$ is at $t=0$.
We thus obtain $h^{\mathrm{NR}}(t, q_i)$ for $i=1,2,3$.
For each waveform, we then compute the amplitude
$A^{\mathrm{NR}}(t, q_i) = |h^{\mathrm{NR}}(t, q_i)|$ and instantaneous
frequency $\omega^{\mathrm{NR}}(t,q_i)=d/dt \arg(h^{\mathrm{NR}}(t,q_i))$,
and interpolate them to a common uniform time grid $t \in [-100M,80M]$
with spacing $0.4 M$, resulting in 450 sample points.  At each time,
we construct a 2nd order interpolating function in $q$ for $A$ and
$\omega$ across the mass ratios $q_i$.  This set of 2x450 interpolants
constitutes the model.

To create a circular merger waveform at arbitrary
$(q,\phi_0,t_{\mathrm{peak}})$, we evaluate these interpolants at each
sample time with the desired $q$, integrate the resulting $\omega$
numerically to get $\phi$, choosing an appropriate integration
constant, then compute the strain $h$ from $A$ and $\phi$.  This
constitutes the Circular Merger Model (CMM).

\begin{figure*}
  \includegraphics{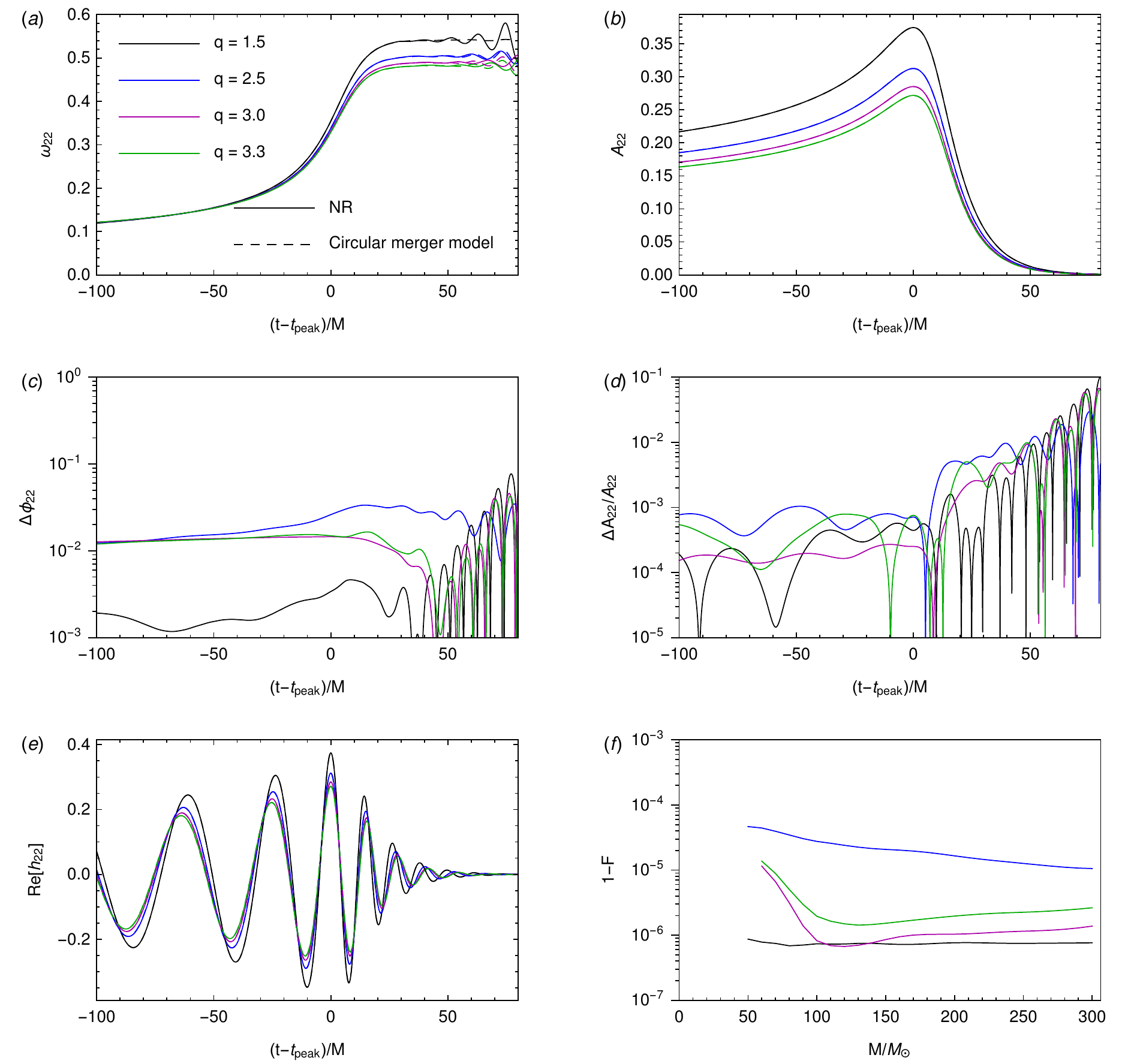}\\
  \caption{Tests of the circular merger model (CMM) for NR waveforms
    which were not used to construct it.  Plotted in (a) and (b) are
    the GW frequency and amplitude from NR simulations (solid curves)
    and the CMM (dashed curves).  We see good agreement.  In (c) and
    (d) are plotted the phase and relative amplitude differences
    between NR and the CMM.  (e) shows the real part of the
    $\ell=2,m=2$ mode of the gravitational wave strain, and (f) shows
    the unfaithfulness between NR and the CMM (see
    Sec.~\ref{sec:faithfulness}).}
  \label{fig:circ_model_tests}
\end{figure*}

To test the CMM, we use additional SXS catalog waveforms with mass
ratios $q = 1.5, 2.5, 3.0, 3.3, 4$ (also used in
\cite{Blackman:2015pia}).  Fig.~\ref{fig:circ_model_tests} shows a
comparison between the CMM and each of the test waveforms.  The solid
lines are the NR data, and the dashed lines are the CMM. We see that
in all cases, $A$ and $\omega$ are visually indistinguishable between
the NR and CMM results, except for some oscillations in $\omega$ at
$t>t_{\mathrm{peak}}+50 M$.  The phase and amplitude differences between
NR and the CMM are also plotted, and we see that the maximum phase
error in the CMM is $\sim 0.15$ radians, and the maximum amplitude
error is $\sim 10\%$ at late times, but only $\sim 3\%$ if the low amplitude
portion at the end of the ringdown is excluded\footnote{In fact, the test waveforms have higher numerical truncation
error than the input waveforms, and this error is comparable to the
differences between the CMM and the test waveforms.}
.

In order to evaluate the faithfulness of the CMM with the NR
waveforms, we have combined the inspiral of an NR test waveform with
the merger from the CMM, blended using a transition function $\mathcal{T}$
(see Eq.~\ref{eqn:plancktaper}) in the
region $t-\tpeak\in[-100M,-80M]$.  As shown in
Fig.~\ref{fig:circ_model_tests}, the unfaithfulness in each case is $1
- F < 4\times 10^{-4}$.

\section{Construction of an IMR waveform model}
\label{sec:constructimr}

\subsection{Motivation and approach}

We have described the eccentric PN model for the inspiral, and a circular model for
the merger, and have shown that the merger from NR is essentially circular.
We now define a method for combining the PN
inspiral (``$x$-model'') with the circular merger model (CMM) based on
a simple blending of the two models in a transition region.  We will
evaluate afterwards how well this has worked.

The starting point is the reference time, at which $x=\xref$.  We
choose the parameters $\eccparamsnot{}$ at this time, $t_{\mathrm{ref}}$.  We
compute the waveform from PN for $t < t_{\mathrm{ref}}$.  We use the CMM
for $t > t_{\mathrm{peak}} - \tcircval{} M$, since we
have shown that a circular model is good after $\tcircval{} M$ before the peak.
Note that we do not yet know the time at which the peak occurs, given
the time of the reference point, $\Delta t = t_{\mathrm{peak}} -
t_{\mathrm{ref}}$.

If the PN waveform agreed well with the NR waveform for $t < \tcirc$,
the model would now be complete, because we
could match the circular waveform frequency and phase to the PN
frequency and phase at $t = \tcirc$.
Unfortunately, the PN waveform cannot be extended reliably up to
$\tcirc$, and it disagrees with the circular NR waveform between $\tref$
and $\tcirc$, so this procedure would result in a merger waveform with a
noticeable error in the time and phase of the peak.
This is not surprising, because the PN approximation is not expected
to be good so close to the merger.

Instead, we adopt a simple model for the time to merger $\Delta t$ and
fit it to the NR simulations.  This model works very well, and
essentially guarantees that the final IMR model will have the waveform
peak at the correct time, to within the errors in $\Delta t$.  Once we
have $\Delta t$, we blend the eccentric PN waveform with the circular
NR interpolated model between $\tref$ and $\tcirc$.  There will be a
discrepancy between the model and NR in this region, and the validity
of this model will be
assessed in Sec.~\ref{sec:modellingresults}.

\subsection{Time to merger}
\label{sec:time_to_merger}

We now determine $\Delta t = t_{\mathrm{peak}}-t_{\mathrm{ref}}$ given
the parameters at $t_{\mathrm{ref}}$.  The most general functional form
would be
\begin{equation}
  {\Delta t}(q,e,l) = \sum_{ijk=0}^\infty a_{ijk} q^i e^j \cos(k l-\alpha_{ijk})
\end{equation}
where we use a Taylor expansion in $q$ and $e$, and a Fourier
series in $l$, since $l$ is a periodic variable.  In order to match
the NR data, we find that we require quadratic terms in $q$ and $e$,
but only the first mode in $l$.  Since there can be no variation with
$l$ when $e = 0$, we must have $a_{i01} = 0$.  The resulting model for $\Delta t$ is
\begin{eqnarray}
  {\Delta t}(q,e,l) &=& \Delta t_0 + a_1 e + a_2 e^2 + b_1 q + b_2 q^2 + \nonumber \\
  && c_1 e \cos(l + c_2) + c_3 e q
\end{eqnarray}
\begin{figure}
  \includegraphics{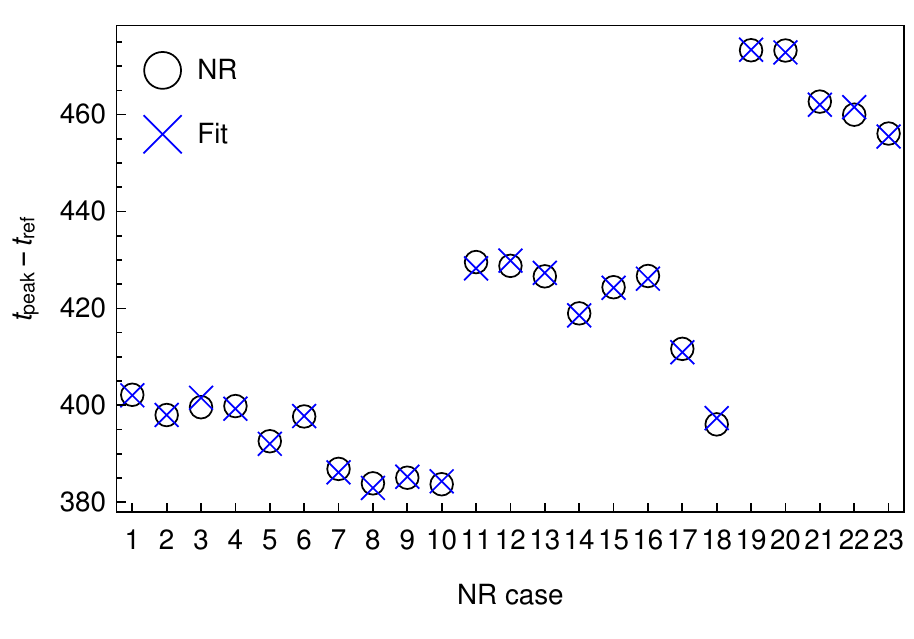}
  \caption{Agreement of the time-to-merger model with NR.  The time
    between the reference point and the peak of $|h_{22}|$ is shown as
    a circle for each NR simulation, and a cross for the fitted model
    from Eq.~\ref{eqn:time_to_merger}.}
  \label{fig:time_to_merger}
\end{figure}
There are 8 unknown parameters, and the model is fitted to all \numsims{} simulations.  
The fitted function is
\begin{eqnarray}
  {\Delta t}(q,e,l) &=& 391.196 + 3.13391 e - 2492.95 e^2 + \nonumber \\
  && 2.77212 q - 17.92 e q + 8.11842 q^2 + \nonumber \\ 
  && 76.4944 e \cos(0.626653 + l) \label{eqn:time_to_merger}
\end{eqnarray}

Fig.~\ref{fig:time_to_merger} shows $\Delta t$ for each NR simulation,
along with the value obtained from the fit. The fit residual is
less than $\pm 1 M$, and the essential functional dependence
of $\Delta t(q,e,l)$ has been captured by the model.  We conclude that
the time of the peak can be predicted from the parameters at $\tref$ to
within $\pm 1 M$.  

\subsection{Combining all the ingredients}

Given the eccentric parameters $\eccparamsnot{}$ at $t_{\mathrm{ref}}$, we
now construct a full IMR waveform.  The eccentric PN waveform is
$h^{\mathrm{PN}}(t)$, such that its parameters at $t = t_{\mathrm{ref}}$
match the desired model parameters.  The circular merger waveform is
$h_{\mathrm{circ}}(t)$, such that the peak occurs at $t=0$.  The waveform is decomposed
into amplitude, $A$, and frequency, $\omega$, as
$h = A e^{i \phi}$, and $\dot \phi = \omega$.  The IMR waveform is
given by
  \begin{eqnarray}
    t_{\mathrm{peak}} &=& t_{\mathrm{ref}} + \Delta t \\
    t_{\mathrm{circ}} &=& t_{\mathrm{peak}} - \tcircval{} M\\
    t_{\mathrm{blend}} &=& t |_{x=\xblendsym}\\
    \alpha(t) &=& \mathcal{T}(t;t_{\mathrm{blend}},t_{\mathrm{circ}}) \\
    A(t) &=& \alpha(t) A^{\mathrm{PN}} + (1-\alpha(t)) A^{\mathrm{circ}}(t-t_{\mathrm{peak}}) \\
    \omega(t) &=& \alpha(t) \omega^{\mathrm{PN}} + (1-\alpha(t)) \omega^{\mathrm{circ}}(t-t_{\mathrm{peak}}) \\
    \phi(t) &=& \int^t \omega(t') dt' \\
    h(t) &=& A(t) e^{i \phi(t)}
  \end{eqnarray}
The start of the blending region is chosen as $\xblendsym =
\xblend$, and the reference point as $\xrefsym = \xref$.
In words, we time-shift the circular waveform so that its peak is in
the correct place according to the time-to-merger fit of
Sec.~\ref{sec:time_to_merger}, and blend the amplitude and frequency of
the PN and circular waveforms using a transition function $\mathcal{T}$
(see Eq.~\ref{eqn:plancktaper}) between
$t_{\mathrm{blend}}$ and $t_{\mathrm{circ}}$ to ensure a smooth transition
in these quantities.  The phase is then computed by integrating the
frequency, leading to the final waveform.

\begin{figure}[h]
  \includegraphics{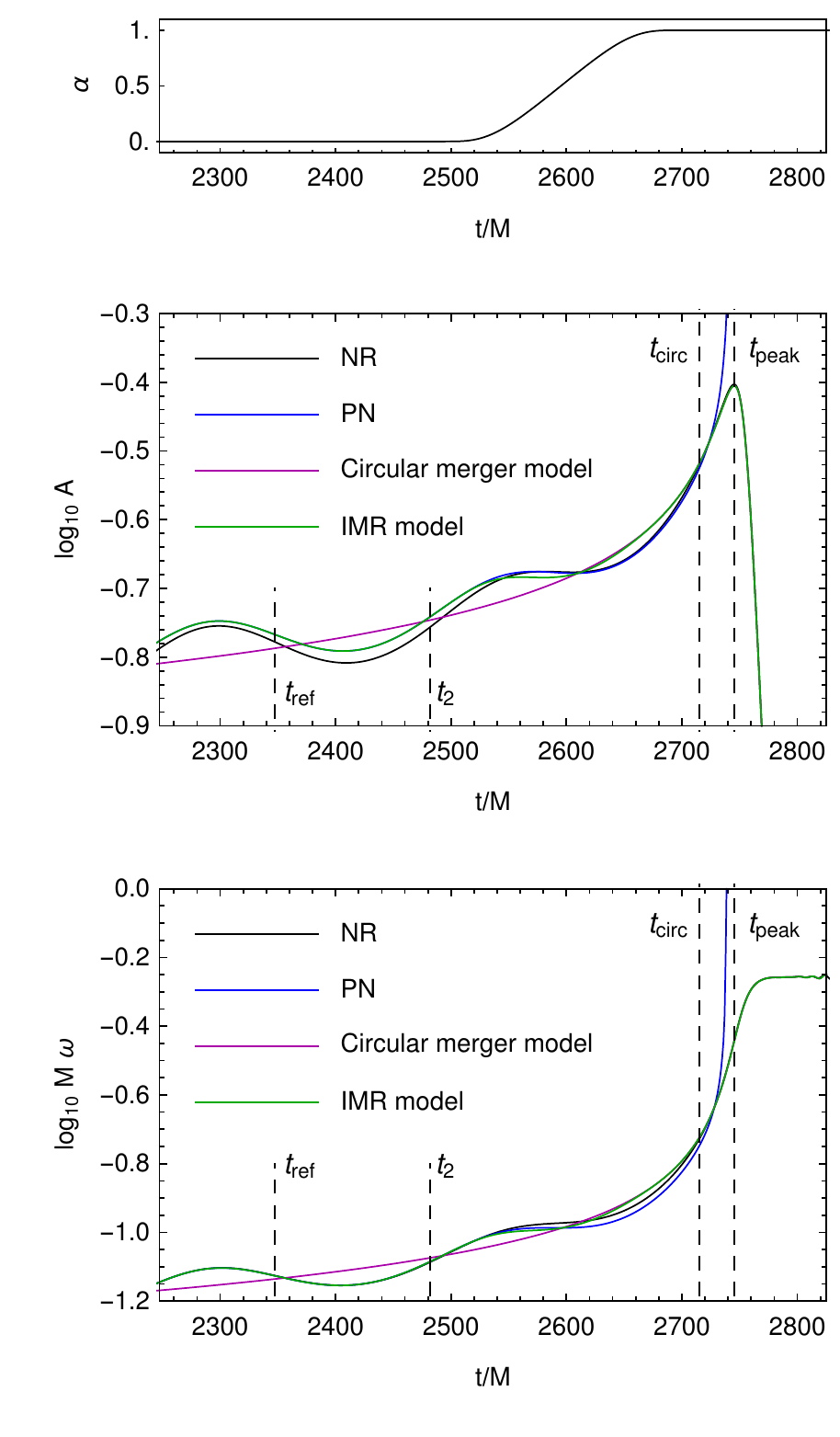}
  \caption{Combining the different ingredients to produce the IMR
    waveform.  The top panel shows the transition function $\alpha$
    which is used to blend the amplitude, $A$, and frequency,
    $\omega$, between the eccentric PN and circular NR waveforms.  The
    middle and bottom panels show $A$ and $\omega$ from the NR
    simulation, the PN model fitted to it at the reference point
    $\tref$, and the circular merger model (CMM) with peak
    $t_{\text{peak}}$ determined from the time-to-merger fit $\Delta T$.
    Between $\tref$ and $t_{\text{peak}}$, the IMR model is constructed
    by blending the PN and CMM quantities using $\alpha$.}
  \label{fig:imr_construction}
\end{figure}

This procedure is illustrated for \caselabel{16} in
Fig.~\ref{fig:imr_construction}, which shows the amplitudes and
frequencies from PN and the circular model, as well as the transition
region in which they are blended.  The NR waveform is shown
for comparison, but no information from the NR waveform (other
than the set of fit parameters at $\tref{}$) is used
in computing the model waveform.

We see in Fig.~\ref{fig:imr_construction} that the IMR waveform
$\omega$ agrees with PN and NR before $\tref{}$, and with the CMM
after $\tcirc{}$.  There is a visible discrepancy between the IMR and
NR frequency between $\tref{}$ and $\tcirc{}$, though
this is small.  The PN waveform breaks down after $\tcirc{}$.  The IMR
amplitude $A$ has a visible disagreement with the NR amplitude,
presumably due to the fact that zeroth-order PN (restricted) waveform
amplitudes are used in the model.

\section{Fourier domain comparisons of waveforms}
\label{sec:fouriercompare}

In this section, we discuss the comparison of waveforms from the point
of view of gravitational wave data analysis, which requires Fourier
representations of the waveforms.  We investigate the effect of
eccentricity on the procedure used to ensure that Fourier transforms
of time-domain truncated waveforms are reliable.

\subsection{Faithfulness}
\label{sec:faithfulness}

The data from a gravitational wave detector is analyzed by a process
of matched filtering against a set of template waveforms in the
frequency domain.  In Sec.~\ref{sec:faithfulnessresults}, we will determine
how well the eccentric IMR waveform model defined in
Sec.~\ref{sec:constructimr} agrees with potential astrophysical sources.
Given two waveforms $h_1(t)$ and $h_2(t)$, their noise-weighted
\textit{overlap} is defined as~\cite{Finn:1992wt}
\begin{equation}
  ( h_1 | h_2 ) \equiv 4\, \textrm{Re} \int_{f_{\text{min}}}^{f_{\text{max}}}
  \frac{\tilde h_1(f) \tilde h^*_2(f)}{ S_n(f) } {\textrm{d}} f \,,
  \label{eq:overlap}
\end{equation}
where $\tilde h_{1,2}(f)$ are the Fourier transforms of the waveforms
and $S_n(f)$ is the one-sided power spectral density (PSD) of the
detector noise.

We examine two Advanced LIGO detector configurations.  
The first, \aligoOOne{}, is representative of the sensitivity of LIGO
during its first observing run.  The noise PSD~\cite{O1PSD} is the one
which was used to place templates for the O1 search, as described in
\cite{Usman:2015kfa}, and we restrict to a frequency range
$f_{\text{min}}=30\Hz$, $f_{\text{max}}=2050\Hz$
The second configuration, \textit{aLIGO design}, is representative of
the sensitivity expected for the final design configuration of
Advanced LIGO.  The noise PSD is the zero-detuned-high-power variant
from \cite{Shoemaker:2010}, with a frequency range $f_{\text{min}} = 10\Hz$,
$f_{\text{max}} = 8192\Hz$.

The \emph{faithfulness} between two waveforms is then defined as the
overlap between the normalized waveforms maximized over relative time
and phase shifts
\begin{equation}
  F = 
\max_{\phi_c, t_c} \frac{
    \left(h_1 (\phi_c, t_c)~|~h_2\right)}{\sqrt{ (h_1|h_1) (h_2|h_2)}} \,.
  \label{eq:faithfulness}
\end{equation}
The faithfulness measures how similar the waveforms would appear to a
gravitational wave detector when the data is analyzed using matched
filtering.

\subsection{Fourier transforms of eccentric NR waveforms}

Computation of the faithfulness, Eq.~\ref{eq:faithfulness}, requires the
Fourier transforms of the
waveforms $h_1$ and $h_2$, which will correspond to the IMR model
waveform and the ``true'' astrophysical waveform, which we take to be
the NR waveform.  Hence, we need to compute the Fourier transforms of
these waveforms.

To estimate the continuum Fourier transform, we use a discrete Fourier
transform (DFT) over the available NR time interval (see,
e.g.~\cite{hansen:fourier}). To minimize Gibbs' phenomena due to
time-domain truncation, the waveform is tapered by multiplying it by a
variant of the Planck taper function~\cite{McKechan:2010kp},
\begin{equation}
\mathcal{T}(t;t_1,t_2) =
\begin{cases}
  0 & \text{for } t \le t_1 \\
  \left [ \exp \left ( \frac{t_2-t_1}{t-t_1} + \frac{t_2-t_1}{t-t_2} \right ) + 1 \right ]^{-1} & \text{for } t_1 < t < t_2 \\
  1 & \text{for } t \ge t_2
\end{cases} \label{eqn:plancktaper}
\end{equation}
at both the start and end of the waveform.  Specifically,
\begin{align}
h_{\ell m}(t) \to {} & h_{\ell m}(t) \\
 & {}\times \mathcal{T}(t;\trel,\trel+250\,M) \\
 & {}\times (1-\mathcal{T}(t;\tpeak+60\,M,\tpeak+80\,M)) \,.
\end{align}
$\trel$ is the \textit{relaxed} time, after which the effects of non-astrophysical junk
radiation in the waveform can be neglected (here chosen as $500\,M$ from
the start of the waveform), and $\tpeak$ is the time of the peak in
$|h_{22}|$, roughly corresponding to the merger.  The waveform is also
resampled to a time step of $0.4 \, M$ (higher frequency content is not important here)
and padded with zeros before
computing the discrete Fourier transform to ensure a sufficiently
small time step in frequency space.

In the quasi-circular case, $\omega_{22}(t) \approx 2 \, \dot \phi$,
where $\omega_{22}(t)$ is the frequency of the dominant instantaneous
GW emission from the binary, and $\dot \phi$ is the orbital angular
velocity.  $\dot \phi$ increases monotonically on the
radiation-reaction timescale.  Intuitively, $\tilde h(\omega)$
consists of contributions from times when $\dot \phi \approx
\omega/2$.  The amplitude of the Fourier transform is $|\tilde h_{22}|
\sim (M \omega)^{-7/6}$ to leading PN order; i.e.~it decreases with
increasing $\omega$ because the binary spends more time, and hence
there is more total GW emission, at lower frequency than at high
frequency, and the increase in the amplitude of emission per orbit at
high frequency is not enough to dominate over this effect.

A quasi-circular NR simulation starts with a given orbital
angular velocity $\dot \phi_0$, and contributions to $\tilde h(\omega)$
for $\omega < 2 \, \dot \phi_0$, which would be present in a real
astrophysical waveform, are not present in the NR waveform.  In other
words, $\tilde
h(\omega)$ for the time-truncated waveform is \textit{unphysical}
below a certain frequency, and its amplitude is strongly suppressed
for $\omega < 2 \, \dot \phi_0$.  Hence, there is a peak at
$\omega_{\text{peak}} \approx 2 \, \dot \phi_0$ in the Fourier transform of the truncated
waveform. Typically, $\tilde h(\omega)$ is found to be relatively free
of Gibbs' phenomena and agrees with longer waveforms for
$\omega > 1.2 \, \omega_{\text{peak}}$ \cite{taracchini:fpeak}.

In the eccentric case, there is no longer a single frequency emitted
at a given time, and $\omega_{22} \approx 2 \, \dot \phi$ oscillates on
the \textit{orbital} timescale (see Eq.~\ref{eq:phidotofu}), so it is
not clear that the minimum frequency at which $\tilde h(\omega)$ is
reliable can be determined using the same criterion in the eccentric
case as in the circular case.

In order to assess the effect of time truncation, we have run one
simulation, \caselabel{54}, starting from a lower orbital
frequency than the others, giving 40 cycles rather than
the typical \typicalcycles{}, and $\sim 6000 M$ of evolution time,
rather than the typical $2500 M$.

\begin{figure}[h]
  \includegraphics{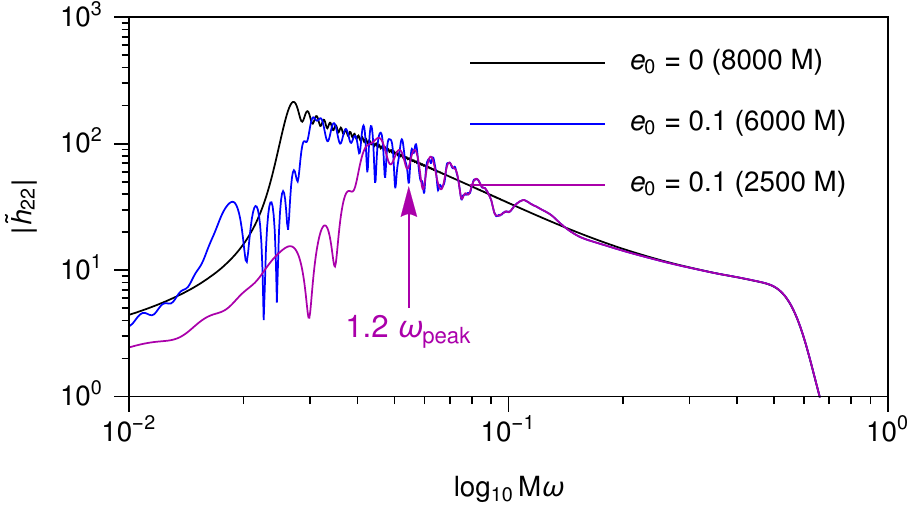}
  \caption{Fourier transforms of circular and eccentric waveforms. The
    eccentric waveform has been truncated in the time domain to two
    different lengths.  This allows us to assess the effect of time-domain
    truncation on the Fourier transform.}
  \label{fig:eccentric_spectrum}
\end{figure}

Fig.~\ref{fig:eccentric_spectrum} shows the amplitudes of the Fourier
transforms of waveforms with eccentricities 0 and 0.1 (\caselabel{39}
and \caselabel{54}). The eccentric waveform is plotted twice; once truncated
in the time domain $6000 M$ before merger, and once truncated $2500 M$
before merger.  We see that at high frequency
($M \omega > 2 \times 10^{-1}$), the effect of eccentricity on the
waveform is negligible, indicating that the merger waveform is more or
less independent of eccentricity.  At intermediate frequency
($4\times 10^{-2} < M \omega < 2 \times 10^{-1}$), we see the
oscillations in $|\tilde h_{22}|$ for $e = 0.1$ characteristic of
eccentricity.  The $e = 0.1$ waveform appears independent of time
truncation for $M \omega \gtrapprox 5 \times 10^{-2}$.  The peaks in
the amplitudes of $\tilde h$ are visible between $2\times 10^{-2}$ and
$\sim 10^{-1}$, depending on the configuration, and for $\omega <
\omega_{\text{peak}}$ in each case, the amplitude drops rapidly to 0 as
$\omega \to 0$.  In this case, it appears that $\tilde h(\omega)$ is
independent of time truncation for $\omega > 1.2 \,
\omega_{\text{peak}}$, as in the circular case, and we assume this in
the analysis that follows.  This is only a preliminary check of the
effect of time truncation on eccentric waveforms, and a more detailed
study in future would be beneficial.

To ensure that the integral in Eq.~\ref{eq:overlap} only covers the
physical part of the waveform $f > 1.2f_{\text{peak}}$, we restrict to
computing the unfaithfulness for systems for which $1.2f_{\text{peak}} <
f_{\text{min}}$.  $f_{\text{peak}}$ scales inversely with the total mass
of the system, so it is only possible to compute the unfaithfulness
for systems with total mass $M > M_{\text{min}}$, where $M_{\text{min}}$
depends on both the length of the NR waveform, and the particular GW
detector considered.  If longer NR waveforms starting from lower
frequency were available, the unfaithfulness could be computed for
lower mass systems.

\section{Modeling results}
\label{sec:modellingresults}

We have described how to generate an eccentric IMR waveform
for a given $e_0$ and $l_0$.  This model will now be tested by comparing the
waveforms from the IMR model to NR.  The parameters of the PN
waveforms used in the comparison are obtained by fitting
$\omega^{\text{PN}}(e_{\text{ref}},l_{\text{ref}})$ to $\omega^{\text{NR}}$ in
a one period window centered on at $x=\xrefsym=\xref$ as described in
Sec.~\ref{sec:measuringecc}.  The relative residual for this fit is
less than $\omfitrefrelres{}\%$ in all cases.

\subsection{Instantaneous gravitational wave frequency}

\begin{figure}[h]
  \includegraphics{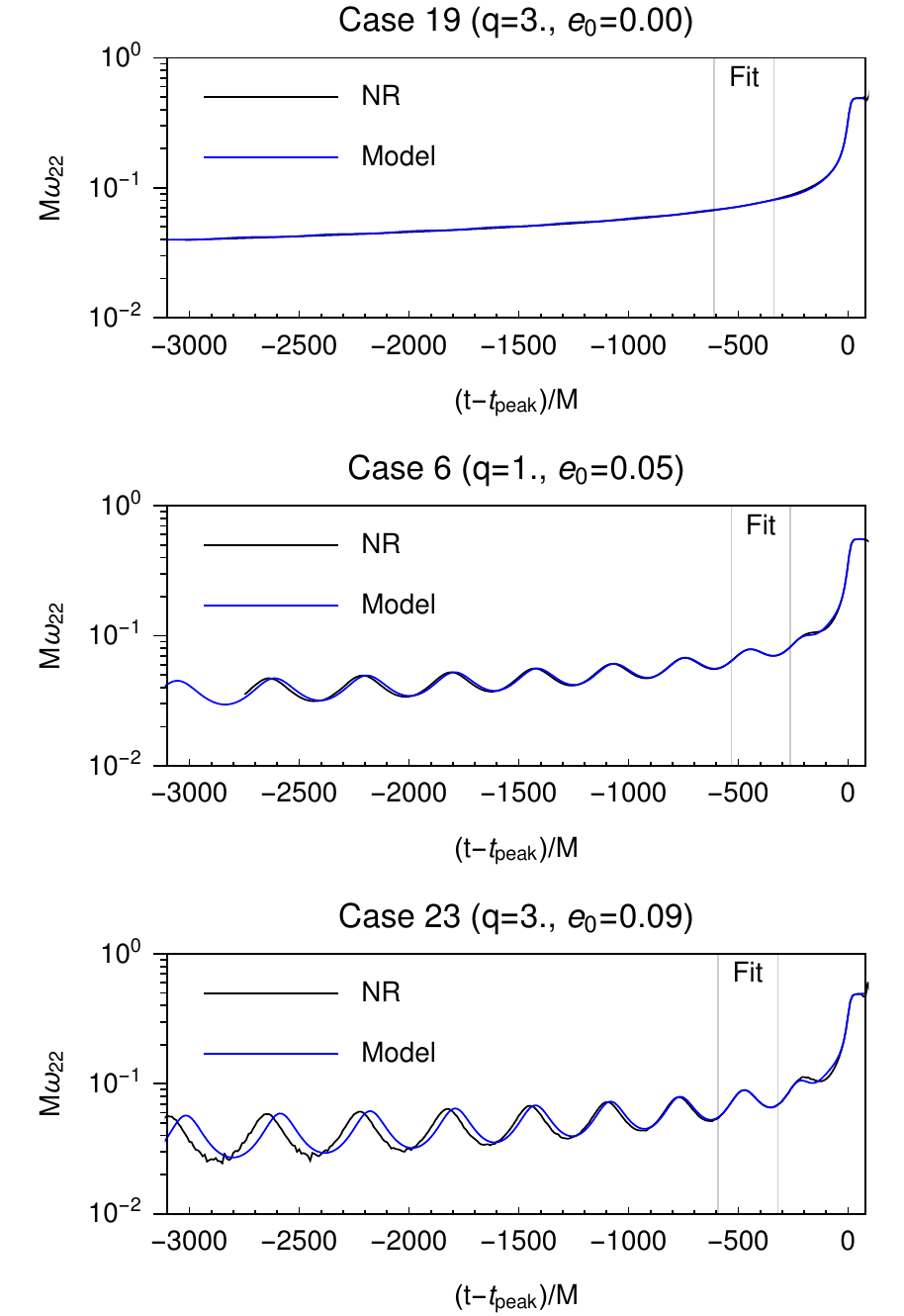}
  \caption{Gravitational wave frequency comparison between IMR model and NR}
  \label{fig:nr_model_frequency}
\end{figure}

Since the PN parameters of the NR waveform are determined by fitting
the instantaneous GW frequency $\omega_{22}$, this quantity is
expected to agree the best between NR and the model, at least within
the fitting window.  Fig.~\ref{fig:nr_model_frequency} shows
$\omega_{22}$ as a function of time for three of the NR simulations.
In each case, the fit window is highlighted.  The top panel shows Case
53 ($q=3$, $e=0$), representing the quasi-circular limit of the
model. We see that at the highest mass ratio studied here, 
$\omega_{22}$ from NR and the model agree well for the duration of the NR
waveform. The middle panel shows \caselabel{16} ($q=1$, $e=0.05$).  For this
equal-mass case with moderate eccentricity, the phase and amplitude of
the oscillations in the NR $\omega_{22}$ are reproduced well by the
model, though there is some dephasing seen at early times.  This
shows that the radiation reaction in the model does not perfectly
capture the evolution of the advance of pericenter, $\Delta \phi$.
The bottom panel shows \caselabel{52}
($q=3$, $e = 0.09$).  This is the most extreme configuration studied, with the highest eccentricity and mass ratio, and shows the limitations of the model.  The dephasing in $\omega_{22}$ at early times is clearly visible, as is the error between the fit window and the merger.

These three cases are examples which broadly represent the performance
of the model across the whole set of NR configurations.  We conclude
that the model reproduces the NR $\omega_{22}$ well, but the
agreement, especially at early times, becomes worse with increasing
mass ratio and eccentricity.

\subsection{Strain}

\begin{figure}[h]
  \includegraphics{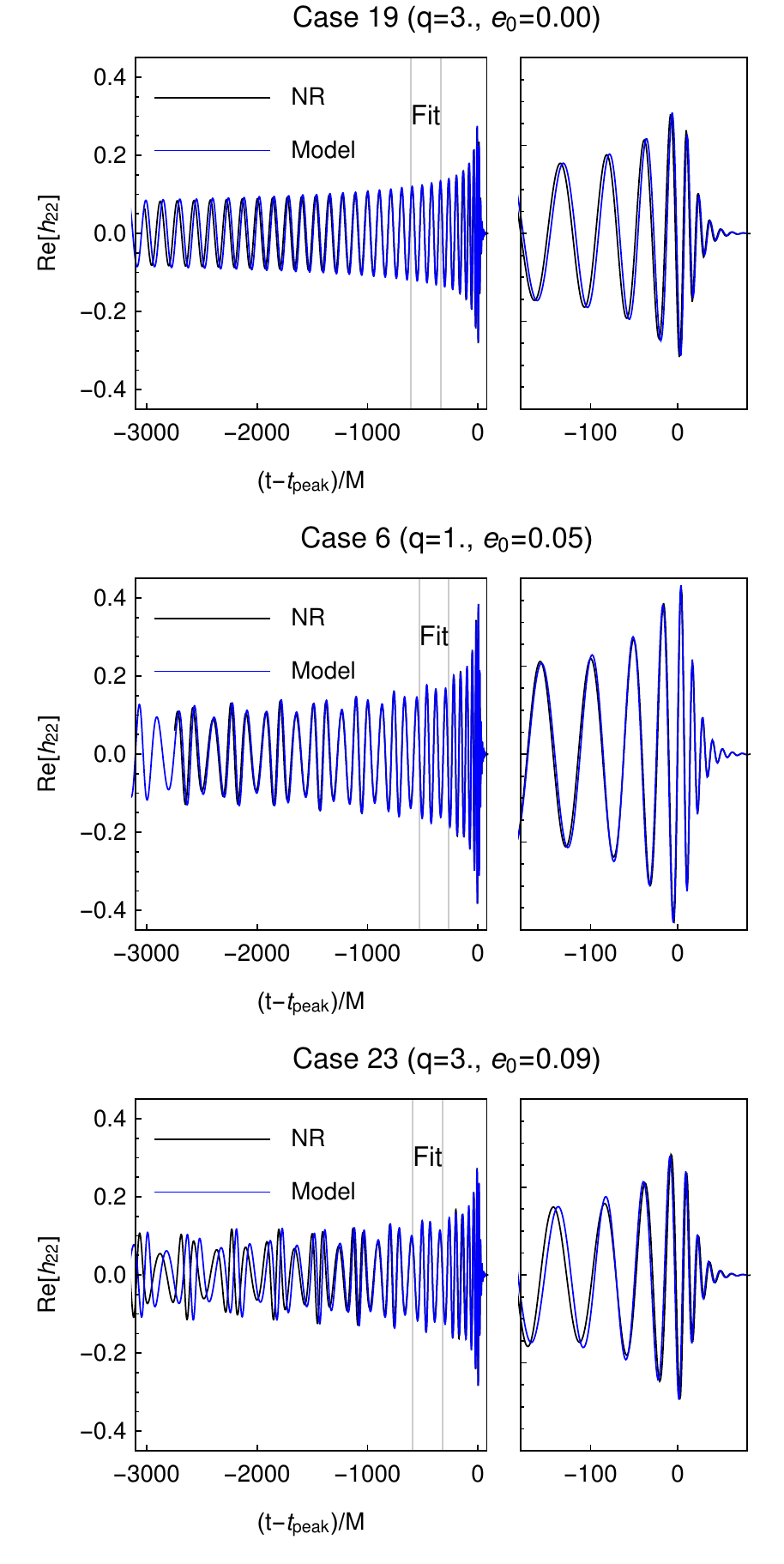}
  \caption{Gravitational wave strain comparison between IMR model and
    NR.  Top: quasi-circular case with $q=3$; some dephasing is
    visible.  Middle: an eccentric equal-mass case, which shows
    excellent agreement.  Bottom: $q=3$ with the highest eccentricity
    configuration, which shows the worst agreement of all the cases.}
  \label{fig:nr_model_strain}
\end{figure}

Fig.~\ref{fig:nr_model_strain} shows the real part of the strain, $\re{} [h_{22}]$ from both
NR and the model.  The right panels highlight the merger and
ringdown, and while the agreement is not perfect, the
model largely agrees with NR.  There is some dephasing visible at
early times for the quasi-circular $q=3$ case. The equal-mass case
with moderate eccentricity agrees very well with NR.  As for the frequency,
for the case
with the highest mass ratio and eccentricity, there is noticeable
dephasing at early times between the model and NR.  Again, this may be improved by
using 3 PN radiation reaction terms in the model.

\subsection{Phase}

\begin{figure}[h]
  \includegraphics{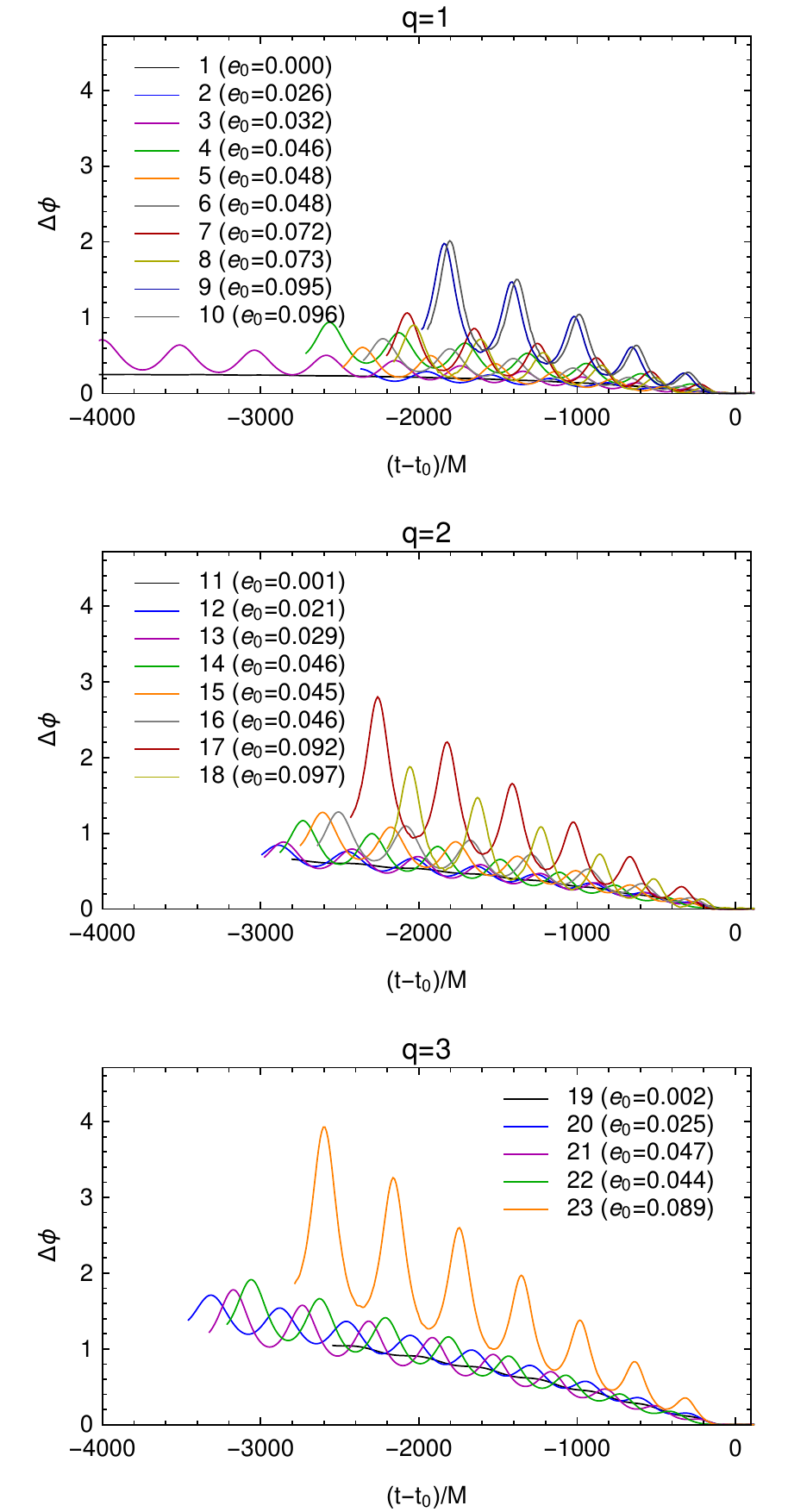}
  \caption{Phase difference between IMR model and NR}
  \label{fig:nr_model_phase_differences}
\end{figure}

Fig.~\ref{fig:nr_model_phase_differences} shows the phase error in the model waveform; $\Delta \phi = \arg h_{22}^{\text{NR}} - \arg h_{22}^{\text{model}}$.
For most of the waveforms, the phase error of the circular case 
gives a lower bound on the error of the eccentric cases.  For $e \lesssim 0.05$,
the phase error
oscillates between the circular value and a value a few times larger.
There is no appreciable effect of eccentricity on the secular growth of the phase error
for these
eccentricities, suggesting that the effect of eccentricity on the
error in the adiabatic evolution is negligible.  For higher
eccentricities, this is no longer the case, and eccentricity appears
to increase the secular phase error.

We expect that adding higher order radiation reaction terms to the
model, as in \cite{Huerta:2016rwp}, will decrease the
phase error.

\subsection{Amplitude}

\begin{figure}[h]
  \includegraphics{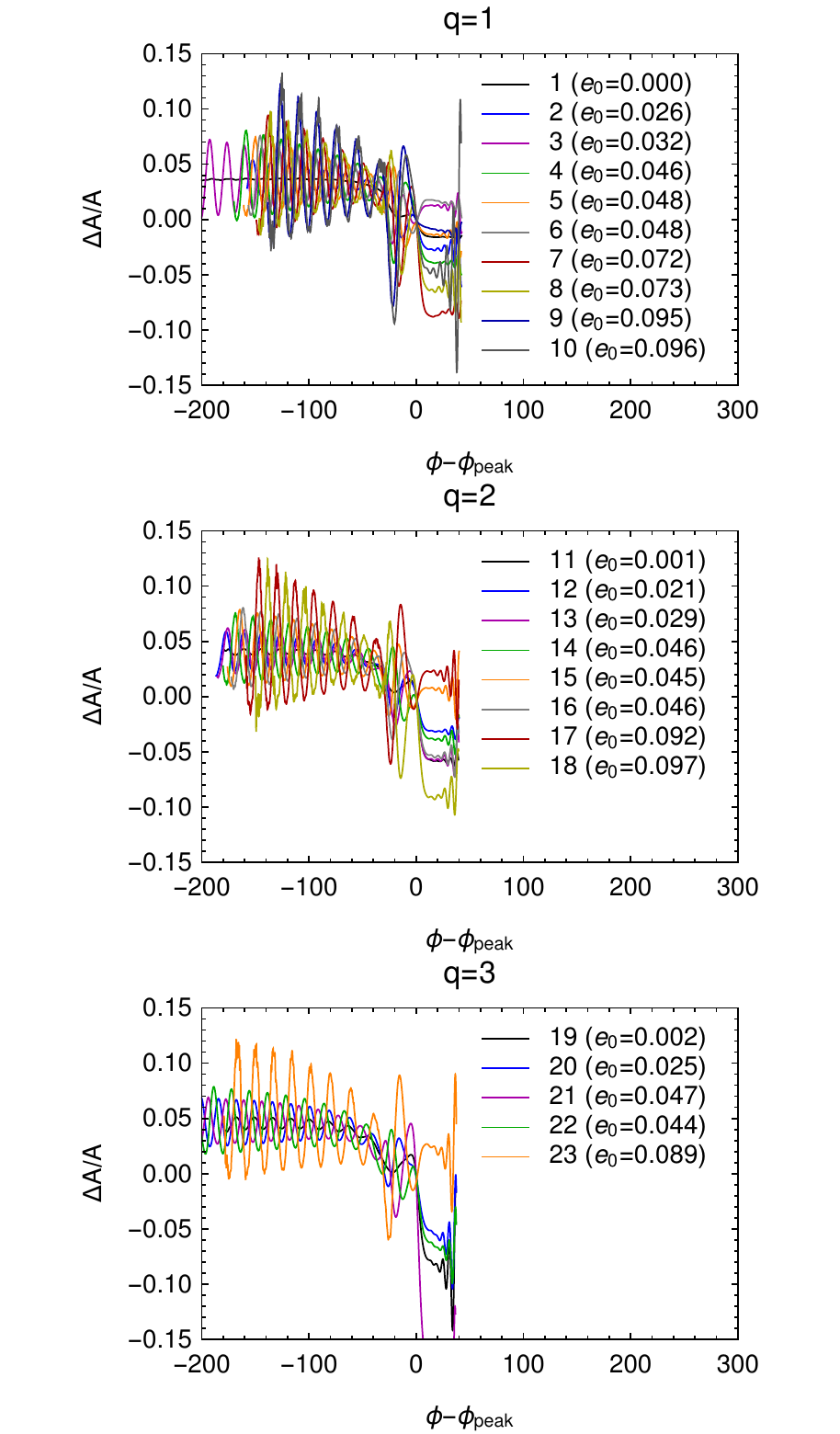}
  \caption{Amplitude difference between IMR model and
    NR}
  \label{fig:nr_model_amp_differences}
\end{figure}

Fig.~\ref{fig:nr_model_amp_differences} shows the relative difference
in the amplitude, $A = |h_{22}|$, between the NR and model waveforms.
Note that we plot $A(\phi)$ instead of $A(t)$, so that phase and
amplitude errors are decoupled.  The amplitude error varies between
4\% and 13\%.

The IMR model incorporates 0 PN restricted waveforms; i.e.~the
expression for the waveform in terms of the orbital quantities,
Eq.~\ref{eqn:h22}, is given by the quadrupole formula. State-of-the-art
quasi-circular models use waveforms which are 2.5 PN accurate.  In the
quasi-circular case, it has been shown (e.g.~in \cite{Boyle2007}) that the use of lower order
waveforms primarily affects the amplitude rather than the phase, so it
is not surprising to see relatively large amplitude error here, even
in cases where the phase errors are fairly small.

\subsection{Faithfulness}
\label{sec:faithfulnessresults}

\begin{figure}[h]
  \includegraphics{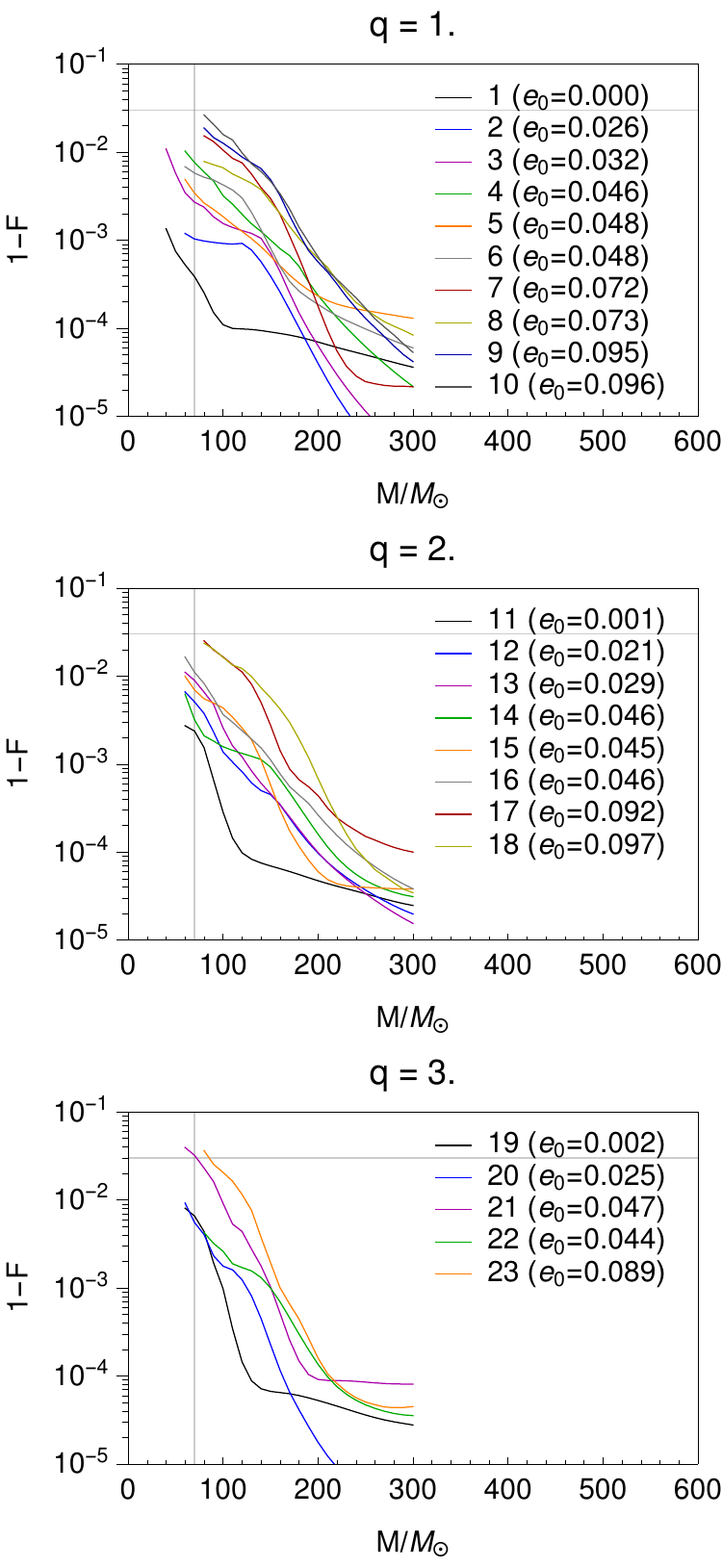}
  \caption{The unfaithfulness, $1-F$, for the \aligoOOne{}
    detector configuration, between the eccentric IMR model
    and the NR simulations, as a function of the total binary mass.
    Masses for which the NR waveform
    starts at a frequency higher than the detector's $f_{\text{min}}$
    are omitted from the plot.  The horizontal line shows the 3\%
    unfaithfulness target, and the vertical line shows $70 \MSun$,
    roughly corresponding to the mass of GW150914.}
  \label{fig:nr_model_unfaith_o1}
\end{figure}

\begin{figure}[h]
  \includegraphics{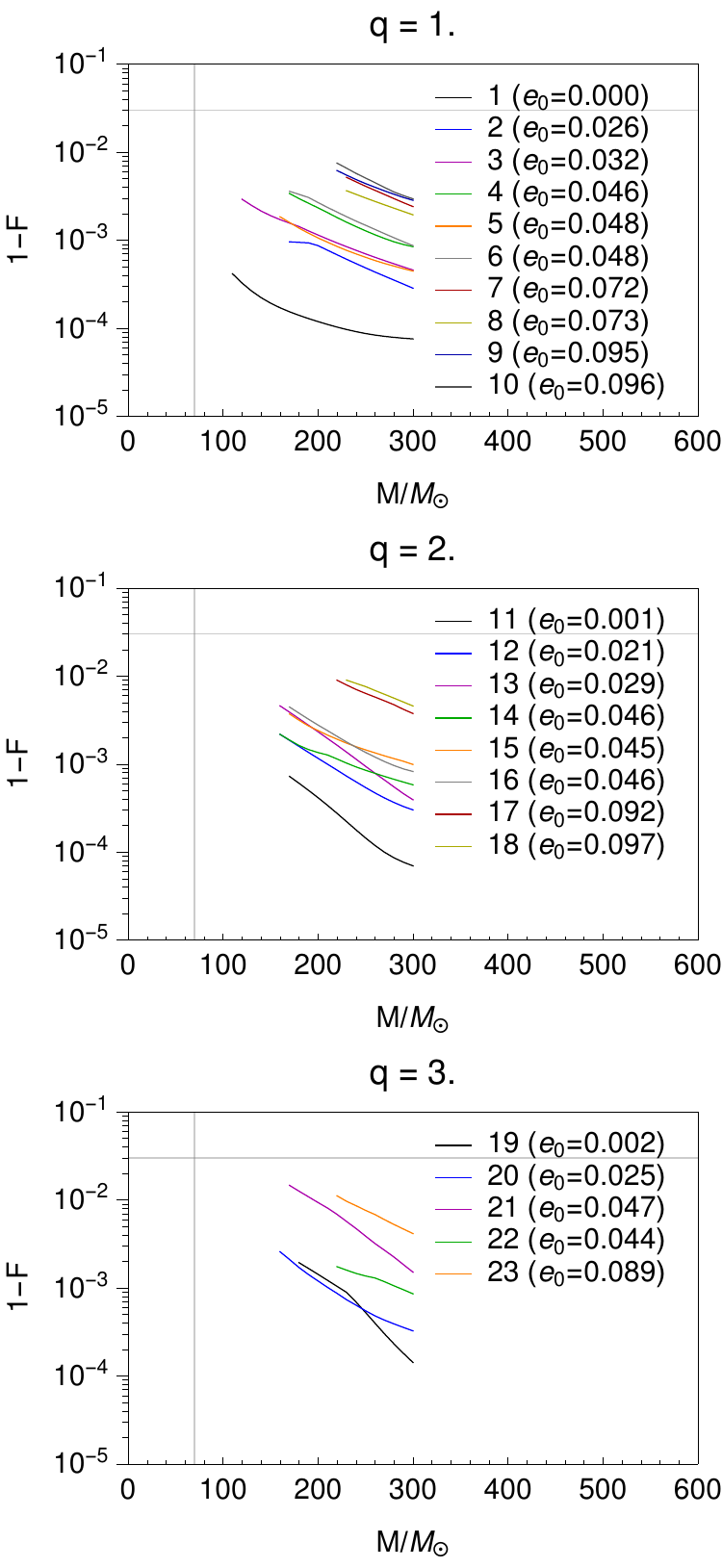}
  \caption{The unfaithfulness, $1-F$, for the
    \aligoDesign{} detector configuration, between the eccentric IMR model
    and the NR simulations, as a function of the total binary mass.
    Masses for which the NR waveform
    starts at a frequency higher than the detector's $f_{\text{min}}$
    are omitted from the plot.  The horizontal line shows the 3\%
    unfaithfulness target, and the vertical line shows $70 \MSun$,
    roughly corresponding to the mass of GW150914.}
  \label{fig:nr_model_unfaith_design}
\end{figure}

Figs.~\ref{fig:nr_model_unfaith_o1} and \ref{fig:nr_model_unfaith_design} show,
for each mass ratio $q$, the
unfaithfulness between the model and each NR waveform for the two advanced LIGO
detector configurations \aligoOOne{} and \aligoDesign{} (see Sec.~\ref{sec:faithfulness}).
The unfaithfulness is plotted only for
the source masses for which the entire NR waveform is in the sensitive
frequency band of the detector.  The unfaithfulness gets
higher as either eccentricity or mass ratio increases.  

For \aligoOOne{},
with a low frequency cutoff of $30 \Hz$, the model has
an unfaithfulness of
less than 3\% for $q=1$ across the entire mass range covered by the NR
waveforms, or $M \ge \minnrmassearly \MSun$.
For $q=3$, the highest eccentricity waveforms have
unfaithfulness less than 3\% only for slightly higher masses, namely $M > 90
\MSun$.
By extrapolating the results in Fig.~\ref{fig:nr_model_unfaith_o1} to
lower mass, it appears that the unfaithfulness of the highest
eccentricity waveforms would probably exceed the $3 \%$ target for
masses $\lesssim 70 \MSun$.

For \aligoDesign{}, with a low frequency cutoff of $10 \Hz$, the model has
faithfulness 3\% for all mass ratios
and eccentricities for which the NR waveform is entirely in band,
however most of the NR waveforms are too short to compute
unfaithfulness for $M \lesssim 180 \MSun$.  In general, for a given
total mass, the unfaithfulness with \aligoDesign{} is greater than
with \aligoOOne{}, so it is reasonable to expect that at masses
$\approx 70 \MSun$, the highest eccentricity waveforms would also
exceed the unfaithfulness target of $3 \%$.

Longer NR waveforms, reaching lower frequencies (for which
the entire waveform is in the sensitive band of the detector for lower
mass systems) will be necessary to accurately assess the performance
of the model for lower mass systems with \aligoDesign{}. Improvements to
the model would be needed to reach acceptable
levels of faithfulness for $M \sim 70 \MSun$ (corresponding to GW150914)
for high eccentricities,
with the maximum usable eccentricity being lower for \aligoDesign{}
than for \aligoOOne{}.

\subsection{Comparison with quasi-circular models}
\label{sec:compareqc}

We now compare the eccentric IMR model with existing quasi-circular
($e=0$) models currently used by
LIGO for estimating the parameters of gravitational wave sources.

Two such models are SEOBNRv4 \cite{Bohe:2016gbl} and IMRPhenomD
\cite{Husa:2015iqa,Khan:2015jqa}.  These models have been compared
with a large number of quasi-circular NR waveforms, and the
unfaithfulness is found to be $< 1\,\%$ for \aligoDesign{}
in almost all cases.  For \aligoOOne{},
which is less sensitive at all frequencies, the
unfaithfulness will be even lower.  The analogues of
Fig.~\ref{fig:nr_model_unfaith_design} are Fig.~2 of \cite{Bohe:2016gbl} and
Fig.~15 of \cite{Khan:2015jqa}.

In the quasi-circular ($e=0$) case, the new eccentric model presented
here has an unfaithfulness $< 1\,\%$ for \aligoOOne{} for $M > 60
\MSun$ (corresponding to the minimum mass for which we are able to
compute unfaithfulness given the length of the NR waveforms), but the
unfaithfulness is larger for larger eccentricities.

The quasi-circular models incorporate higher order PN
radiation reaction, as well as additional features designed to
increase the accuracy of the dynamics, whereas the eccentric model
uses simple PN for the inspiral.  Further, the quasi-circular models
have been tested and calibrated against NR waveforms with much lower
initial frequencies.  As a result, we expect the low frequency
behavior of the quasi-circular models to be superior to the eccentric
model in the quasi-circular limit.

However, as we have shown, the eccentric model is faithful to the NR
data for the last $\sim 20$ cycles before the merger, when the
eccentricity is $\lesssim 0.1$ at a time $\approx
\typicalcyclesrefpeak{}$ cycles before the merger.

\section{Conclusions}
\label{sec:conclusions}


We have presented \numsims{} new publicly-available non-spinning NR
BBH simulations with initial eccentricities ranging from 0 to 0.2 and
mass ratios from 1 to 3, including the merger and the preceding
\typicalcycles{} gravitational wave cycles.  When considered as
sources for gravitational wave detectors, the NR waveforms start below
30\,\Hz{} for systems of total mass $M>\minnrmassearly \MSun$, and below 10\,\Hz{} for
systems of total mass $M > \minnrmassdesign \MSun$.

We have demonstrated that the circularisation of eccentric binary
black hole systems in the last few cycles before the merger first
reported in \cite{Hinder:2007qu} for equal-mass systems extends to systems with mass
ratio up to $q = 3$.

We have shown that an existing PN model for the inspiral can be fitted
to the NR data over one radial period shortly before the merger, and
have quantified how the error in the PN model grows at earlier times.
The results depend on mass ratio and eccentricity, with higher mass
ratios and eccentricities generally showing a larger error at early times. 

For all the NR waveforms, the PN model remains accurate to within about half
a gravitational wave cycle across the entire NR waveform, but this is
unlikely to be the case for longer NR waveforms.

Using the fact that the NR waveforms circularize shortly before the
merger, we have shown that the merger can be represented using a circular
model.  A simple circular model was built from a small number of
circular NR waveforms by interpolating them in the mass ratio $q$.  A
full IMR model was then constructed by blending the
PN inspiral model with the NR-interpolated circular model.  The
combination relies on knowing the time $\Delta t(q,e,l)$ between a
reference point ($x=\xref{}$) in the PN waveform and the peak of the
merger waveform, and we have derived an accurate empirical fitting formula
for $\Delta t$ from the NR waveforms.

We have compared the IMR model with all the NR simulations.  For
$f_{\text{low}} = 30 \Hz$, and a detector configuration, \aligoOOne{}, corresponding to
the first observing run of advanced LIGO, the eccentric model
has a faithfulness of $\ge
97\%$ with the corresponding NR waveform for systems of total mass
$M\ge\maxmtarget{}M_{\odot}$ for all the NR simulations
($e_{\text{ref}}\le\maxecc{}$, $q\le\maxq{}$), and for systems of
total mass $M\ge\lowmass{}M_{\odot}$, the faithfulness is over
$\targetfaith{}\%$ for $e_{\text{ref}}\lesssim\lowmassecc{}$ and
$q\le\lowmassq{}$.
%


The availability of eccentric IMR waveform models such as the model
presented in \cite{Huerta:2016rwp}, and the model presented here,
which has been calibrated to and validated against NR simulations, is
the first step towards measuring the eccentricity of binary black hole
mergers through their gravitational wave emission.  We have shown that
the merger can be accurately represented by a simple combination of
eccentric PN and circular NR results.

Note, however, that the present model has been validated only for the
last $\sim \typicalcycles$ cycles before the merger, corresponding to the finite
length of the NR simulations used.  For systems of sufficiently high
mass that this is the only part of the waveform which is in the
sensitive band of the detector, for example sources similar to
GW150914, the model may be useful for parameter estimation.
For systems where longer waveforms
are required, i.e.~lower mass systems for which both the merger and
more of the early inspiral is in the sensitive band of the detector,
the model is probably not sufficiently faithful to the general
relativistic waveform for reliable results to be obtained.

The model has been calibrated to NR simulations with parameters in the
range ($e_{\text{ref}}\le\maxecc{}$, $q\le\maxq{}$), but it can be
evaluated outside its range of calibration.  The model is not a
small-$e$ expansion, so in principle it may be evaluated for any
$e<1$.  However, the PN approximation, which is an expansion in
$v/c$, will break down for high eccentricities if the velocity becomes
too large at pericenter.  The circular merger model, as described
here, can only be reliably evaluated within its calibration range $q\le4$,
but by including more NR simulations, an extension to higher
mass ratio would be straightforward.

The simulations and model presented here are restricted to the case of
non-spinning binaries.  For interesting applications to gravitational
wave data, the model will need to be extended to include the effects
of spin, otherwise it's possible that the effects of eccentricity
and spin could be confused.  We also model only the dominant $\ell=2,m=\pm2$ spherical
harmonic modes.  While the effects of sub-dominant modes are likely
more important for eccentric systems than for circular systems, we
expect the effects to be small for the moderate eccentricities studied
here. Finally, while our model is fully
3~PN accurate in the conservative dynamics, the radiation reaction
terms are implemented only up to 2~PN, in contrast to the model of
\cite{Huerta:2016rwp} which is 3~PN in both the conservative and
radiative effects, and also contains improvements for high mass ratios
based on the test-mass limit.  We expect that the performance of our
model during the early inspiral (when aligned just before the merger)
would be improved with these modifications, but we leave that to
future work.

\begin{acknowledgments}
We thank Alessandra Buonanno, Tim Dietrich, Sergei Ossokine and
Michael P\"urrer for helpful discussions.  Calculations were performed
using the Spectral Einstein code ({\tt SpEC})~\cite{SpECwebsite} on
the Datura and Minerva clusters at the AEI.
Analysis of the numerical data was performed using SimulationTools for
Mathematica~\cite{SimulationToolsWeb}.
L.K. gratefully
acknowledges support from the Sherman Fairchild Foundation and NSF
grants PHY-1606654 and AST-1333129 at Cornell.  H.P. gratefully
acknowledges funding from NSERC of Canada, the Ontario Early
Researcher Awards Program, the Canada Research Chairs Program, and the
Canadian Institute for Advanced Research.

\end{acknowledgments}

\bibliography{references}

\begin{thebibliography}{77}%
\makeatletter
\providecommand \@ifxundefined [1]{%
 \@ifx{#1\undefined}
}%
\providecommand \@ifnum [1]{%
 \ifnum #1\expandafter \@firstoftwo
 \else \expandafter \@secondoftwo
 \fi
}%
\providecommand \@ifx [1]{%
 \ifx #1\expandafter \@firstoftwo
 \else \expandafter \@secondoftwo
 \fi
}%
\providecommand \natexlab [1]{#1}%
\providecommand \enquote  [1]{``#1''}%
\providecommand \bibnamefont  [1]{#1}%
\providecommand \bibfnamefont [1]{#1}%
\providecommand \citenamefont [1]{#1}%
\providecommand \href@noop [0]{\@secondoftwo}%
\providecommand \href [0]{\begingroup \@sanitize@url \@href}%
\providecommand \@href[1]{\@@startlink{#1}\@@href}%
\providecommand \@@href[1]{\endgroup#1\@@endlink}%
\providecommand \@sanitize@url [0]{\catcode `\\12\catcode `\$12\catcode
  `\&12\catcode `\#12\catcode `\^12\catcode `\_12\catcode `\%12\relax}%
\providecommand \@@startlink[1]{}%
\providecommand \@@endlink[0]{}%
\providecommand \url  [0]{\begingroup\@sanitize@url \@url }%
\providecommand \@url [1]{\endgroup\@href {#1}{\urlprefix }}%
\providecommand \urlprefix  [0]{URL }%
\providecommand \Eprint [0]{\href }%
\providecommand \doibase [0]{http://dx.doi.org/}%
\providecommand \selectlanguage [0]{\@gobble}%
\providecommand \bibinfo  [0]{\@secondoftwo}%
\providecommand \bibfield  [0]{\@secondoftwo}%
\providecommand \translation [1]{[#1]}%
\providecommand \BibitemOpen [0]{}%
\providecommand \bibitemStop [0]{}%
\providecommand \bibitemNoStop [0]{.\EOS\space}%
\providecommand \EOS [0]{\spacefactor3000\relax}%
\providecommand \BibitemShut  [1]{\csname bibitem#1\endcsname}%
\let\auto@bib@innerbib\@empty
\bibitem [{\citenamefont {Abbott}\ \emph
  {et~al.}(2016{\natexlab{a}})\citenamefont {Abbott} \emph
  {et~al.}}]{Abbott:2016blz}%
  \BibitemOpen
  \bibfield  {author} {\bibinfo {author} {\bibfnamefont {B.~P.}\ \bibnamefont
  {Abbott}} \emph {et~al.} (\bibinfo {collaboration} {Virgo, LIGO
  Scientific}),\ }\href {\doibase 10.1103/PhysRevLett.116.061102} {\bibfield
  {journal} {\bibinfo  {journal} {Phys. Rev. Lett.}\ }\textbf {\bibinfo
  {volume} {116}},\ \bibinfo {pages} {061102} (\bibinfo {year}
  {2016}{\natexlab{a}})},\ \Eprint {http://arxiv.org/abs/1602.03837}
  {arXiv:1602.03837 [gr-qc]} \BibitemShut {NoStop}%
\bibitem [{\citenamefont {Abbott}\ \emph
  {et~al.}(2016{\natexlab{b}})\citenamefont {Abbott} \emph
  {et~al.}}]{Abbott:2016nmj}%
  \BibitemOpen
  \bibfield  {author} {\bibinfo {author} {\bibfnamefont {B.~P.}\ \bibnamefont
  {Abbott}} \emph {et~al.} (\bibinfo {collaboration} {Virgo, LIGO
  Scientific}),\ }\href {\doibase 10.1103/PhysRevLett.116.241103} {\bibfield
  {journal} {\bibinfo  {journal} {Phys. Rev. Lett.}\ }\textbf {\bibinfo
  {volume} {116}},\ \bibinfo {pages} {241103} (\bibinfo {year}
  {2016}{\natexlab{b}})},\ \Eprint {http://arxiv.org/abs/1606.04855}
  {arXiv:1606.04855 [gr-qc]} \BibitemShut {NoStop}%
\bibitem [{\citenamefont {Abbott}\ \emph {et~al.}(2017)\citenamefont {Abbott}
  \emph {et~al.}}]{Abbott:2017vtc}%
  \BibitemOpen
  \bibfield  {author} {\bibinfo {author} {\bibfnamefont {B.~P.}\ \bibnamefont
  {Abbott}} \emph {et~al.} (\bibinfo {collaboration} {VIRGO, LIGO
  Scientific}),\ }\href {\doibase 10.1103/PhysRevLett.118.221101} {\bibfield
  {journal} {\bibinfo  {journal} {Phys. Rev. Lett.}\ }\textbf {\bibinfo
  {volume} {118}},\ \bibinfo {pages} {221101} (\bibinfo {year} {2017})},\
  \Eprint {http://arxiv.org/abs/1706.01812} {arXiv:1706.01812 [gr-qc]}
  \BibitemShut {NoStop}%
\bibitem [{\citenamefont {Taracchini}\ \emph {et~al.}(2014)\citenamefont
  {Taracchini} \emph {et~al.}}]{Taracchini:2013rva}%
  \BibitemOpen
  \bibfield  {author} {\bibinfo {author} {\bibfnamefont {A.}~\bibnamefont
  {Taracchini}} \emph {et~al.},\ }\href {\doibase 10.1103/PhysRevD.89.061502}
  {\bibfield  {journal} {\bibinfo  {journal} {Phys. Rev.}\ }\textbf {\bibinfo
  {volume} {D89}},\ \bibinfo {pages} {061502} (\bibinfo {year} {2014})},\
  \Eprint {http://arxiv.org/abs/1311.2544} {arXiv:1311.2544 [gr-qc]}
  \BibitemShut {NoStop}%
\bibitem [{\citenamefont {Khan}\ \emph {et~al.}(2016)\citenamefont {Khan},
  \citenamefont {Husa}, \citenamefont {Hannam}, \citenamefont {Ohme},
  \citenamefont {Pürrer}, \citenamefont {Jiménez~Forteza},\ and\
  \citenamefont {Bohé}}]{Khan:2015jqa}%
  \BibitemOpen
  \bibfield  {author} {\bibinfo {author} {\bibfnamefont {S.}~\bibnamefont
  {Khan}}, \bibinfo {author} {\bibfnamefont {S.}~\bibnamefont {Husa}}, \bibinfo
  {author} {\bibfnamefont {M.}~\bibnamefont {Hannam}}, \bibinfo {author}
  {\bibfnamefont {F.}~\bibnamefont {Ohme}}, \bibinfo {author} {\bibfnamefont
  {M.}~\bibnamefont {Pürrer}}, \bibinfo {author} {\bibfnamefont
  {X.}~\bibnamefont {Jiménez~Forteza}}, \ and\ \bibinfo {author}
  {\bibfnamefont {A.}~\bibnamefont {Bohé}},\ }\href {\doibase
  10.1103/PhysRevD.93.044007} {\bibfield  {journal} {\bibinfo  {journal} {Phys.
  Rev.}\ }\textbf {\bibinfo {volume} {D93}},\ \bibinfo {pages} {044007}
  (\bibinfo {year} {2016})},\ \Eprint {http://arxiv.org/abs/1508.07253}
  {arXiv:1508.07253 [gr-qc]} \BibitemShut {NoStop}%
\bibitem [{\citenamefont {Mroue}\ \emph {et~al.}(2013)\citenamefont {Mroue}
  \emph {et~al.}}]{Mroue:2013xna}%
  \BibitemOpen
  \bibfield  {author} {\bibinfo {author} {\bibfnamefont {A.~H.}\ \bibnamefont
  {Mroue}} \emph {et~al.},\ }\href {\doibase 10.1103/PhysRevLett.111.241104}
  {\bibfield  {journal} {\bibinfo  {journal} {Phys. Rev. Lett.}\ }\textbf
  {\bibinfo {volume} {111}},\ \bibinfo {pages} {241104} (\bibinfo {year}
  {2013})},\ \Eprint {http://arxiv.org/abs/1304.6077} {arXiv:1304.6077 [gr-qc]}
  \BibitemShut {NoStop}%
\bibitem [{\citenamefont {Abbott}\ \emph
  {et~al.}(2016{\natexlab{c}})\citenamefont {Abbott} \emph
  {et~al.}}]{TheLIGOScientific:2016wfe}%
  \BibitemOpen
  \bibfield  {author} {\bibinfo {author} {\bibfnamefont {B.~P.}\ \bibnamefont
  {Abbott}} \emph {et~al.} (\bibinfo {collaboration} {Virgo, LIGO
  Scientific}),\ }\href {\doibase 10.1103/PhysRevLett.116.241102} {\bibfield
  {journal} {\bibinfo  {journal} {Phys. Rev. Lett.}\ }\textbf {\bibinfo
  {volume} {116}},\ \bibinfo {pages} {241102} (\bibinfo {year}
  {2016}{\natexlab{c}})},\ \Eprint {http://arxiv.org/abs/1602.03840}
  {arXiv:1602.03840 [gr-qc]} \BibitemShut {NoStop}%
\bibitem [{\citenamefont {Peters}(1964)}]{Peters:1964zz}%
  \BibitemOpen
  \bibfield  {author} {\bibinfo {author} {\bibfnamefont {P.~C.}\ \bibnamefont
  {Peters}},\ }\href {\doibase 10.1103/PhysRev.136.B1224} {\bibfield  {journal}
  {\bibinfo  {journal} {Phys. Rev.}\ }\textbf {\bibinfo {volume} {136}},\
  \bibinfo {pages} {B1224} (\bibinfo {year} {1964})}\BibitemShut {NoStop}%
\bibitem [{\citenamefont {{Wen}}(2003)}]{2003ApJ...598..419W}%
  \BibitemOpen
  \bibfield  {author} {\bibinfo {author} {\bibfnamefont {L.}~\bibnamefont
  {{Wen}}},\ }\href {\doibase 10.1086/378794} {\bibfield  {journal} {\bibinfo
  {journal} {Astrophys. J.}\ }\textbf {\bibinfo {volume} {598}},\ \bibinfo
  {pages} {419} (\bibinfo {year} {2003})}\BibitemShut {NoStop}%
\bibitem [{\citenamefont {Benacquista}(2002)}]{Benacquista:2002kf}%
  \BibitemOpen
  \bibfield  {author} {\bibinfo {author} {\bibfnamefont {M.~J.}\ \bibnamefont
  {Benacquista}},\ }\href@noop {} {\bibfield  {journal} {\bibinfo  {journal}
  {Living Rev. Rel.}\ }\textbf {\bibinfo {volume} {5}},\ \bibinfo {pages} {2}
  (\bibinfo {year} {2002})},\ \Eprint {http://arxiv.org/abs/astro-ph/0202056}
  {arXiv:astro-ph/0202056} \BibitemShut {NoStop}%
\bibitem [{\citenamefont {Gultekin}\ \emph {et~al.}(2004)\citenamefont
  {Gultekin}, \citenamefont {Miller},\ and\ \citenamefont
  {Hamilton}}]{Gultekin:2004pm}%
  \BibitemOpen
  \bibfield  {author} {\bibinfo {author} {\bibfnamefont {K.}~\bibnamefont
  {Gultekin}}, \bibinfo {author} {\bibfnamefont {M.~C.}\ \bibnamefont
  {Miller}}, \ and\ \bibinfo {author} {\bibfnamefont {D.~P.}\ \bibnamefont
  {Hamilton}},\ }\href {\doibase 10.1086/424809} {\bibfield  {journal}
  {\bibinfo  {journal} {Astrophys. J.}\ }\textbf {\bibinfo {volume} {616}},\
  \bibinfo {pages} {221} (\bibinfo {year} {2004})},\ \Eprint
  {http://arxiv.org/abs/astro-ph/0402532} {arXiv:astro-ph/0402532} \BibitemShut
  {NoStop}%
\bibitem [{\citenamefont {{Thorne}}\ and\ \citenamefont
  {{Braginskii}}(1976)}]{1976ApJ...204L...1T}%
  \BibitemOpen
  \bibfield  {author} {\bibinfo {author} {\bibfnamefont {K.~S.}\ \bibnamefont
  {{Thorne}}}\ and\ \bibinfo {author} {\bibfnamefont {V.~B.}\ \bibnamefont
  {{Braginskii}}},\ }\href@noop {} {\bibfield  {journal} {\bibinfo  {journal}
  {Astrophys. J. Lett.}\ }\textbf {\bibinfo {volume} {204}},\ \bibinfo {pages}
  {L1} (\bibinfo {year} {1976})}\BibitemShut {NoStop}%
\bibitem [{\citenamefont {Blaes}\ \emph {et~al.}(2002)\citenamefont {Blaes},
  \citenamefont {Lee},\ and\ \citenamefont {Socrates}}]{Blaes:2002cs}%
  \BibitemOpen
  \bibfield  {author} {\bibinfo {author} {\bibfnamefont {O.}~\bibnamefont
  {Blaes}}, \bibinfo {author} {\bibfnamefont {M.~H.}\ \bibnamefont {Lee}}, \
  and\ \bibinfo {author} {\bibfnamefont {A.}~\bibnamefont {Socrates}},\ }\href
  {\doibase 10.1086/342655} {\bibfield  {journal} {\bibinfo  {journal}
  {Astrophys. J.}\ }\textbf {\bibinfo {volume} {578}},\ \bibinfo {pages} {775}
  (\bibinfo {year} {2002})},\ \Eprint {http://arxiv.org/abs/astro-ph/0203370}
  {arXiv:astro-ph/0203370} \BibitemShut {NoStop}%
\bibitem [{\citenamefont {Dotti}\ \emph {et~al.}(2006)\citenamefont {Dotti},
  \citenamefont {Colpi},\ and\ \citenamefont {Haardt}}]{Dotti:2005kq}%
  \BibitemOpen
  \bibfield  {author} {\bibinfo {author} {\bibfnamefont {M.}~\bibnamefont
  {Dotti}}, \bibinfo {author} {\bibfnamefont {M.}~\bibnamefont {Colpi}}, \ and\
  \bibinfo {author} {\bibfnamefont {F.}~\bibnamefont {Haardt}},\ }\href
  {\doibase 10.1111/j.1365-2966.2005.09956.x} {\bibfield  {journal} {\bibinfo
  {journal} {Mon. Not. Roy. Astron. Soc.}\ }\textbf {\bibinfo {volume} {367}},\
  \bibinfo {pages} {103} (\bibinfo {year} {2006})},\ \Eprint
  {http://arxiv.org/abs/astro-ph/0509813} {arXiv:astro-ph/0509813 [astro-ph]}
  \BibitemShut {NoStop}%
\bibitem [{\citenamefont {Antonini}\ \emph {et~al.}(2016)\citenamefont
  {Antonini}, \citenamefont {Chatterjee}, \citenamefont {Rodriguez},
  \citenamefont {Morscher}, \citenamefont {Pattabiraman}, \citenamefont
  {Kalogera},\ and\ \citenamefont {Rasio}}]{Antonini:2015zsa}%
  \BibitemOpen
  \bibfield  {author} {\bibinfo {author} {\bibfnamefont {F.}~\bibnamefont
  {Antonini}}, \bibinfo {author} {\bibfnamefont {S.}~\bibnamefont
  {Chatterjee}}, \bibinfo {author} {\bibfnamefont {C.~L.}\ \bibnamefont
  {Rodriguez}}, \bibinfo {author} {\bibfnamefont {M.}~\bibnamefont {Morscher}},
  \bibinfo {author} {\bibfnamefont {B.}~\bibnamefont {Pattabiraman}}, \bibinfo
  {author} {\bibfnamefont {V.}~\bibnamefont {Kalogera}}, \ and\ \bibinfo
  {author} {\bibfnamefont {F.~A.}\ \bibnamefont {Rasio}},\ }\href {\doibase
  10.3847/0004-637X/816/2/65} {\bibfield  {journal} {\bibinfo  {journal}
  {Astrophys. J.}\ }\textbf {\bibinfo {volume} {816}},\ \bibinfo {pages} {65}
  (\bibinfo {year} {2016})},\ \Eprint {http://arxiv.org/abs/1509.05080}
  {arXiv:1509.05080 [astro-ph.GA]} \BibitemShut {NoStop}%
\bibitem [{\citenamefont {Rodriguez}\ \emph {et~al.}(2016)\citenamefont
  {Rodriguez}, \citenamefont {Chatterjee},\ and\ \citenamefont
  {Rasio}}]{Rodriguez:2016kxx}%
  \BibitemOpen
  \bibfield  {author} {\bibinfo {author} {\bibfnamefont {C.~L.}\ \bibnamefont
  {Rodriguez}}, \bibinfo {author} {\bibfnamefont {S.}~\bibnamefont
  {Chatterjee}}, \ and\ \bibinfo {author} {\bibfnamefont {F.~A.}\ \bibnamefont
  {Rasio}},\ }\href {\doibase 10.1103/PhysRevD.93.084029} {\bibfield  {journal}
  {\bibinfo  {journal} {Phys. Rev.}\ }\textbf {\bibinfo {volume} {D93}},\
  \bibinfo {pages} {084029} (\bibinfo {year} {2016})},\ \Eprint
  {http://arxiv.org/abs/1602.02444} {arXiv:1602.02444 [astro-ph.HE]}
  \BibitemShut {NoStop}%
\bibitem [{\citenamefont {Maccarone}\ \emph {et~al.}(2007)\citenamefont
  {Maccarone}, \citenamefont {Kundu}, \citenamefont {Zepf},\ and\ \citenamefont
  {Rhode}}]{Maccarone:2007dd}%
  \BibitemOpen
  \bibfield  {author} {\bibinfo {author} {\bibfnamefont {T.~J.}\ \bibnamefont
  {Maccarone}}, \bibinfo {author} {\bibfnamefont {A.}~\bibnamefont {Kundu}},
  \bibinfo {author} {\bibfnamefont {S.~E.}\ \bibnamefont {Zepf}}, \ and\
  \bibinfo {author} {\bibfnamefont {K.~L.}\ \bibnamefont {Rhode}},\ }\href
  {\doibase 10.1038/nature05434} {\bibfield  {journal} {\bibinfo  {journal}
  {Nature}\ }\textbf {\bibinfo {volume} {445}},\ \bibinfo {pages} {183}
  (\bibinfo {year} {2007})},\ \Eprint {http://arxiv.org/abs/astro-ph/0701310}
  {arXiv:astro-ph/0701310 [astro-ph]} \BibitemShut {NoStop}%
\bibitem [{\citenamefont {Strader}\ \emph {et~al.}(2012)\citenamefont
  {Strader}, \citenamefont {Chomiuk}, \citenamefont {Maccarone}, \citenamefont
  {Miller-Jones},\ and\ \citenamefont {Seth}}]{Strader:2012wj}%
  \BibitemOpen
  \bibfield  {author} {\bibinfo {author} {\bibfnamefont {J.}~\bibnamefont
  {Strader}}, \bibinfo {author} {\bibfnamefont {L.}~\bibnamefont {Chomiuk}},
  \bibinfo {author} {\bibfnamefont {T.}~\bibnamefont {Maccarone}}, \bibinfo
  {author} {\bibfnamefont {J.}~\bibnamefont {Miller-Jones}}, \ and\ \bibinfo
  {author} {\bibfnamefont {A.}~\bibnamefont {Seth}},\ }\href {\doibase
  10.1038/nature11490} {\bibfield  {journal} {\bibinfo  {journal} {Nature}\
  }\textbf {\bibinfo {volume} {490}},\ \bibinfo {pages} {71} (\bibinfo {year}
  {2012})},\ \Eprint {http://arxiv.org/abs/1210.0901} {arXiv:1210.0901
  [astro-ph.HE]} \BibitemShut {NoStop}%
\bibitem [{\citenamefont {Chomiuk}\ \emph {et~al.}(2013)\citenamefont
  {Chomiuk}, \citenamefont {Strader}, \citenamefont {Maccarone}, \citenamefont
  {Miller-Jones}, \citenamefont {Heinke}, \citenamefont {Noyola}, \citenamefont
  {Seth},\ and\ \citenamefont {Ransom}}]{Chomiuk:2013qya}%
  \BibitemOpen
  \bibfield  {author} {\bibinfo {author} {\bibfnamefont {L.}~\bibnamefont
  {Chomiuk}}, \bibinfo {author} {\bibfnamefont {J.}~\bibnamefont {Strader}},
  \bibinfo {author} {\bibfnamefont {T.~J.}\ \bibnamefont {Maccarone}}, \bibinfo
  {author} {\bibfnamefont {J.~C.~A.}\ \bibnamefont {Miller-Jones}}, \bibinfo
  {author} {\bibfnamefont {C.}~\bibnamefont {Heinke}}, \bibinfo {author}
  {\bibfnamefont {E.}~\bibnamefont {Noyola}}, \bibinfo {author} {\bibfnamefont
  {A.~C.}\ \bibnamefont {Seth}}, \ and\ \bibinfo {author} {\bibfnamefont
  {S.}~\bibnamefont {Ransom}},\ }\href {\doibase 10.1088/0004-637X/777/1/69}
  {\bibfield  {journal} {\bibinfo  {journal} {Astrophys. J.}\ }\textbf
  {\bibinfo {volume} {777}},\ \bibinfo {pages} {69} (\bibinfo {year} {2013})},\
  \Eprint {http://arxiv.org/abs/1306.6624} {arXiv:1306.6624 [astro-ph.HE]}
  \BibitemShut {NoStop}%
\bibitem [{\citenamefont {Rodriguez}\ \emph {et~al.}(2015)\citenamefont
  {Rodriguez}, \citenamefont {Morscher}, \citenamefont {Pattabiraman},
  \citenamefont {Chatterjee}, \citenamefont {Haster},\ and\ \citenamefont
  {Rasio}}]{Rodriguez:2015oxa}%
  \BibitemOpen
  \bibfield  {author} {\bibinfo {author} {\bibfnamefont {C.~L.}\ \bibnamefont
  {Rodriguez}}, \bibinfo {author} {\bibfnamefont {M.}~\bibnamefont {Morscher}},
  \bibinfo {author} {\bibfnamefont {B.}~\bibnamefont {Pattabiraman}}, \bibinfo
  {author} {\bibfnamefont {S.}~\bibnamefont {Chatterjee}}, \bibinfo {author}
  {\bibfnamefont {C.-J.}\ \bibnamefont {Haster}}, \ and\ \bibinfo {author}
  {\bibfnamefont {F.~A.}\ \bibnamefont {Rasio}},\ }\href {\doibase
  10.1103/PhysRevLett.116.029901, 10.1103/PhysRevLett.115.051101} {\bibfield
  {journal} {\bibinfo  {journal} {Phys. Rev. Lett.}\ }\textbf {\bibinfo
  {volume} {115}},\ \bibinfo {pages} {051101} (\bibinfo {year} {2015})},\
  \bibinfo {note} {[Erratum: Phys. Rev. Lett.116,no.2,029901(2016)]},\ \Eprint
  {http://arxiv.org/abs/1505.00792} {arXiv:1505.00792 [astro-ph.HE]}
  \BibitemShut {NoStop}%
\bibitem [{\citenamefont {Peters}\ and\ \citenamefont
  {Mathews}(1963)}]{PhysRev.131.435}%
  \BibitemOpen
  \bibfield  {author} {\bibinfo {author} {\bibfnamefont {P.~C.}\ \bibnamefont
  {Peters}}\ and\ \bibinfo {author} {\bibfnamefont {J.}~\bibnamefont
  {Mathews}},\ }\href {\doibase 10.1103/PhysRev.131.435} {\bibfield  {journal}
  {\bibinfo  {journal} {Phys. Rev.}\ }\textbf {\bibinfo {volume} {131}},\
  \bibinfo {pages} {435} (\bibinfo {year} {1963})}\BibitemShut {NoStop}%
\bibitem [{\citenamefont {Wagoner}\ and\ \citenamefont
  {Will}(1976)}]{Wagoner:1976am}%
  \BibitemOpen
  \bibfield  {author} {\bibinfo {author} {\bibfnamefont {R.~V.}\ \bibnamefont
  {Wagoner}}\ and\ \bibinfo {author} {\bibfnamefont {C.~M.}\ \bibnamefont
  {Will}},\ }\href {\doibase 10.1086/154886} {\bibfield  {journal} {\bibinfo
  {journal} {Astrophys. J.}\ }\textbf {\bibinfo {volume} {210}},\ \bibinfo
  {pages} {764} (\bibinfo {year} {1976})},\ \bibinfo {note} {[Erratum:
  Astrophys. J.215,984(1977)]}\BibitemShut {NoStop}%
\bibitem [{\citenamefont {Blanchet}\ and\ \citenamefont
  {Sch{\"a}fer}(1989)}]{Blanchet:1989cu}%
  \BibitemOpen
  \bibfield  {author} {\bibinfo {author} {\bibfnamefont {L.}~\bibnamefont
  {Blanchet}}\ and\ \bibinfo {author} {\bibfnamefont {G.}~\bibnamefont
  {Sch{\"a}fer}},\ }\href {\doibase 10.1093/mnras/239.3.845} {\bibfield
  {journal} {\bibinfo  {journal} {Monthly Notices of the Royal Astronomical
  Society}\ }\textbf {\bibinfo {volume} {239}},\ \bibinfo {pages} {845}
  (\bibinfo {year} {1989})}\BibitemShut {NoStop}%
\bibitem [{\citenamefont {{Junker}}\ and\ \citenamefont
  {Sch{\"a}fer}(1992)}]{1992MNRAS.254..146J}%
  \BibitemOpen
  \bibfield  {author} {\bibinfo {author} {\bibfnamefont {W.}~\bibnamefont
  {{Junker}}}\ and\ \bibinfo {author} {\bibfnamefont {G.}~\bibnamefont
  {Sch{\"a}fer}},\ }\href {\doibase 10.1093/mnras/254.1.146} {\bibfield
  {journal} {\bibinfo  {journal} {Monthly Notices of the Royal Astronomical
  Society}\ }\textbf {\bibinfo {volume} {254}},\ \bibinfo {pages} {146}
  (\bibinfo {year} {1992})}\BibitemShut {NoStop}%
\bibitem [{\citenamefont {Blanchet}\ and\ \citenamefont
  {Sch{\"a}fer}(1993)}]{Blanchet:1993ec}%
  \BibitemOpen
  \bibfield  {author} {\bibinfo {author} {\bibfnamefont {L.}~\bibnamefont
  {Blanchet}}\ and\ \bibinfo {author} {\bibfnamefont {G.}~\bibnamefont
  {Sch{\"a}fer}},\ }\href {\doibase 10.1088/0264-9381/10/12/026} {\bibfield
  {journal} {\bibinfo  {journal} {Class. Quant. Grav.}\ }\textbf {\bibinfo
  {volume} {10}},\ \bibinfo {pages} {2699} (\bibinfo {year}
  {1993})}\BibitemShut {NoStop}%
\bibitem [{\citenamefont {Rieth}\ and\ \citenamefont
  {Sch{\"a}fer}(1997)}]{Rieth:1997mk}%
  \BibitemOpen
  \bibfield  {author} {\bibinfo {author} {\bibfnamefont {R.}~\bibnamefont
  {Rieth}}\ and\ \bibinfo {author} {\bibfnamefont {G.}~\bibnamefont
  {Sch{\"a}fer}},\ }\href {\doibase 10.1088/0264-9381/14/8/029} {\bibfield
  {journal} {\bibinfo  {journal} {Class. Quant. Grav.}\ }\textbf {\bibinfo
  {volume} {14}},\ \bibinfo {pages} {2357} (\bibinfo {year}
  {1997})}\BibitemShut {NoStop}%
\bibitem [{\citenamefont {Damour}\ and\ \citenamefont
  {Sch{\"a}fer}(1988)}]{Damour:1988mr}%
  \BibitemOpen
  \bibfield  {author} {\bibinfo {author} {\bibfnamefont {T.}~\bibnamefont
  {Damour}}\ and\ \bibinfo {author} {\bibfnamefont {G.}~\bibnamefont
  {Sch{\"a}fer}},\ }\href {\doibase 10.1007/BF02828697} {\bibfield  {journal}
  {\bibinfo  {journal} {Nuovo Cim.}\ }\textbf {\bibinfo {volume} {B101}},\
  \bibinfo {pages} {127} (\bibinfo {year} {1988})}\BibitemShut {NoStop}%
\bibitem [{\citenamefont {{Sch{\"a}fer}}\ and\ \citenamefont
  {{Wex}}(1993)}]{Schaefer93}%
  \BibitemOpen
  \bibfield  {author} {\bibinfo {author} {\bibfnamefont {G.}~\bibnamefont
  {{Sch{\"a}fer}}}\ and\ \bibinfo {author} {\bibfnamefont {N.}~\bibnamefont
  {{Wex}}},\ }\href {\doibase 10.1016/0375-9601(93)90758-R} {\bibfield
  {journal} {\bibinfo  {journal} {Physics Letters A}\ }\textbf {\bibinfo
  {volume} {174}},\ \bibinfo {pages} {196} (\bibinfo {year}
  {1993})}\BibitemShut {NoStop}%
\bibitem [{\citenamefont {{Wex}}(1995)}]{1995CQGra..12..983W}%
  \BibitemOpen
  \bibfield  {author} {\bibinfo {author} {\bibfnamefont {N.}~\bibnamefont
  {{Wex}}},\ }\href {\doibase 10.1088/0264-9381/12/4/009} {\bibfield  {journal}
  {\bibinfo  {journal} {Class. Quant. Grav.}\ }\textbf {\bibinfo {volume}
  {12}},\ \bibinfo {pages} {983} (\bibinfo {year} {1995})}\BibitemShut
  {NoStop}%
\bibitem [{\citenamefont {Damour}\ \emph {et~al.}(2004)\citenamefont {Damour},
  \citenamefont {Gopakumar},\ and\ \citenamefont {Iyer}}]{Damour:2004bz}%
  \BibitemOpen
  \bibfield  {author} {\bibinfo {author} {\bibfnamefont {T.}~\bibnamefont
  {Damour}}, \bibinfo {author} {\bibfnamefont {A.}~\bibnamefont {Gopakumar}}, \
  and\ \bibinfo {author} {\bibfnamefont {B.~R.}\ \bibnamefont {Iyer}},\ }\href
  {\doibase 10.1103/PhysRevD.70.064028} {\bibfield  {journal} {\bibinfo
  {journal} {Phys. Rev.}\ }\textbf {\bibinfo {volume} {D70}},\ \bibinfo {pages}
  {064028} (\bibinfo {year} {2004})},\ \Eprint
  {http://arxiv.org/abs/gr-qc/0404128} {arXiv:gr-qc/0404128} \BibitemShut
  {NoStop}%
\bibitem [{\citenamefont {{K{\"o}nigsd{\"o}rffer}}\ and\ \citenamefont
  {{Gopakumar}}(2006)}]{2006PhRvD..73l4012K}%
  \BibitemOpen
  \bibfield  {author} {\bibinfo {author} {\bibfnamefont {C.}~\bibnamefont
  {{K{\"o}nigsd{\"o}rffer}}}\ and\ \bibinfo {author} {\bibfnamefont
  {A.}~\bibnamefont {{Gopakumar}}},\ }\href {\doibase
  10.1103/PhysRevD.73.124012} {\bibfield  {journal} {\bibinfo  {journal} {Phys.
  Rev.}\ }\textbf {\bibinfo {volume} {D73}},\ \bibinfo {pages} {124012}
  (\bibinfo {year} {2006})}\BibitemShut {NoStop}%
\bibitem [{\citenamefont {Memmesheimer}\ \emph {et~al.}(2004)\citenamefont
  {Memmesheimer}, \citenamefont {Gopakumar},\ and\ \citenamefont
  {Sch{\"a}fer}}]{Memmesheimer:2004cv}%
  \BibitemOpen
  \bibfield  {author} {\bibinfo {author} {\bibfnamefont {R.-M.}\ \bibnamefont
  {Memmesheimer}}, \bibinfo {author} {\bibfnamefont {A.}~\bibnamefont
  {Gopakumar}}, \ and\ \bibinfo {author} {\bibfnamefont {G.}~\bibnamefont
  {Sch{\"a}fer}},\ }\href {\doibase 10.1103/PhysRevD.70.104011} {\bibfield
  {journal} {\bibinfo  {journal} {Phys. Rev.}\ }\textbf {\bibinfo {volume}
  {D70}},\ \bibinfo {pages} {104011} (\bibinfo {year} {2004})},\ \Eprint
  {http://arxiv.org/abs/gr-qc/0407049} {arXiv:gr-qc/0407049} \BibitemShut
  {NoStop}%
\bibitem [{\citenamefont {{Arun}}\ \emph
  {et~al.}(2008{\natexlab{a}})\citenamefont {{Arun}}, \citenamefont
  {{Blanchet}}, \citenamefont {{Iyer}},\ and\ \citenamefont
  {{Qusailah}}}]{Arun:2007sg}%
  \BibitemOpen
  \bibfield  {author} {\bibinfo {author} {\bibfnamefont {K.~G.}\ \bibnamefont
  {{Arun}}}, \bibinfo {author} {\bibfnamefont {L.}~\bibnamefont {{Blanchet}}},
  \bibinfo {author} {\bibfnamefont {B.~R.}\ \bibnamefont {{Iyer}}}, \ and\
  \bibinfo {author} {\bibfnamefont {M.~S.~S.}\ \bibnamefont {{Qusailah}}},\
  }\href {\doibase 10.1103/PhysRevD.77.064035} {\bibfield  {journal} {\bibinfo
  {journal} {Phys. Rev.}\ }\textbf {\bibinfo {volume} {D77}},\ \bibinfo {pages}
  {064035} (\bibinfo {year} {2008}{\natexlab{a}})},\ \Eprint
  {http://arxiv.org/abs/0711.0302} {arXiv:0711.0302} \BibitemShut {NoStop}%
\bibitem [{\citenamefont {{Arun}}\ \emph
  {et~al.}(2008{\natexlab{b}})\citenamefont {{Arun}}, \citenamefont
  {{Blanchet}}, \citenamefont {{Iyer}},\ and\ \citenamefont
  {{Qusailah}}}]{Arun:2007rg}%
  \BibitemOpen
  \bibfield  {author} {\bibinfo {author} {\bibfnamefont {K.~G.}\ \bibnamefont
  {{Arun}}}, \bibinfo {author} {\bibfnamefont {L.}~\bibnamefont {{Blanchet}}},
  \bibinfo {author} {\bibfnamefont {B.~R.}\ \bibnamefont {{Iyer}}}, \ and\
  \bibinfo {author} {\bibfnamefont {M.~S.~S.}\ \bibnamefont {{Qusailah}}},\
  }\href {\doibase 10.1103/PhysRevD.77.064034} {\bibfield  {journal} {\bibinfo
  {journal} {Phys. Rev.}\ }\textbf {\bibinfo {volume} {D77}},\ \bibinfo {pages}
  {064034} (\bibinfo {year} {2008}{\natexlab{b}})},\ \Eprint
  {http://arxiv.org/abs/0711.0250} {arXiv:0711.0250} \BibitemShut {NoStop}%
\bibitem [{\citenamefont {Arun}\ \emph {et~al.}(2009)\citenamefont {Arun},
  \citenamefont {Blanchet}, \citenamefont {Iyer},\ and\ \citenamefont
  {Sinha}}]{arunblanchetam}%
  \BibitemOpen
  \bibfield  {author} {\bibinfo {author} {\bibfnamefont {K.~G.}\ \bibnamefont
  {Arun}}, \bibinfo {author} {\bibfnamefont {L.}~\bibnamefont {Blanchet}},
  \bibinfo {author} {\bibfnamefont {B.~R.}\ \bibnamefont {Iyer}}, \ and\
  \bibinfo {author} {\bibfnamefont {S.}~\bibnamefont {Sinha}},\ }\href
  {\doibase 10.1103/PhysRevD.80.124018} {\bibfield  {journal} {\bibinfo
  {journal} {Phys. Rev.}\ }\textbf {\bibinfo {volume} {D80}},\ \bibinfo {pages}
  {124018} (\bibinfo {year} {2009})},\ \Eprint {http://arxiv.org/abs/0908.3854}
  {arXiv:0908.3854 [gr-qc]} \BibitemShut {NoStop}%
\bibitem [{\citenamefont {Sperhake}\ \emph {et~al.}(2008)\citenamefont
  {Sperhake} \emph {et~al.}}]{Sperhake:2007gu}%
  \BibitemOpen
  \bibfield  {author} {\bibinfo {author} {\bibfnamefont {U.}~\bibnamefont
  {Sperhake}} \emph {et~al.},\ }\href {\doibase 10.1103/PhysRevD.78.064069}
  {\bibfield  {journal} {\bibinfo  {journal} {Phys. Rev.}\ }\textbf {\bibinfo
  {volume} {D78}},\ \bibinfo {pages} {064069} (\bibinfo {year} {2008})},\
  \Eprint {http://arxiv.org/abs/0710.3823} {arXiv:0710.3823 [gr-qc]}
  \BibitemShut {NoStop}%
\bibitem [{\citenamefont {Hinder}\ \emph {et~al.}(2008)\citenamefont {Hinder},
  \citenamefont {Vaishnav}, \citenamefont {Herrmann}, \citenamefont
  {Shoemaker},\ and\ \citenamefont {Laguna}}]{Hinder:2007qu}%
  \BibitemOpen
  \bibfield  {author} {\bibinfo {author} {\bibfnamefont {I.}~\bibnamefont
  {Hinder}}, \bibinfo {author} {\bibfnamefont {B.}~\bibnamefont {Vaishnav}},
  \bibinfo {author} {\bibfnamefont {F.}~\bibnamefont {Herrmann}}, \bibinfo
  {author} {\bibfnamefont {D.}~\bibnamefont {Shoemaker}}, \ and\ \bibinfo
  {author} {\bibfnamefont {P.}~\bibnamefont {Laguna}},\ }\href {\doibase
  10.1103/PhysRevD.77.081502} {\bibfield  {journal} {\bibinfo  {journal}
  {Phys.Rev.}\ }\textbf {\bibinfo {volume} {D77}},\ \bibinfo {pages} {081502}
  (\bibinfo {year} {2008})},\ \Eprint {http://arxiv.org/abs/0710.5167}
  {arXiv:0710.5167 [gr-qc]} \BibitemShut {NoStop}%
\bibitem [{\citenamefont {Hinder}\ \emph {et~al.}(2010)\citenamefont {Hinder},
  \citenamefont {Herrmann}, \citenamefont {Laguna},\ and\ \citenamefont
  {Shoemaker}}]{Hinder:2008kv}%
  \BibitemOpen
  \bibfield  {author} {\bibinfo {author} {\bibfnamefont {I.}~\bibnamefont
  {Hinder}}, \bibinfo {author} {\bibfnamefont {F.}~\bibnamefont {Herrmann}},
  \bibinfo {author} {\bibfnamefont {P.}~\bibnamefont {Laguna}}, \ and\ \bibinfo
  {author} {\bibfnamefont {D.}~\bibnamefont {Shoemaker}},\ }\href {\doibase
  10.1103/PhysRevD.82.024033} {\bibfield  {journal} {\bibinfo  {journal} {Phys.
  Rev.}\ }\textbf {\bibinfo {volume} {D82}},\ \bibinfo {pages} {024033}
  (\bibinfo {year} {2010})},\ \Eprint {http://arxiv.org/abs/0806.1037}
  {arXiv:0806.1037 [gr-qc]} \BibitemShut {NoStop}%
\bibitem [{\citenamefont {Mroue}\ \emph {et~al.}(2010)\citenamefont {Mroue},
  \citenamefont {Pfeiffer}, \citenamefont {Kidder},\ and\ \citenamefont
  {Teukolsky}}]{Mroue:2010re}%
  \BibitemOpen
  \bibfield  {author} {\bibinfo {author} {\bibfnamefont {A.~H.}\ \bibnamefont
  {Mroue}}, \bibinfo {author} {\bibfnamefont {H.~P.}\ \bibnamefont {Pfeiffer}},
  \bibinfo {author} {\bibfnamefont {L.~E.}\ \bibnamefont {Kidder}}, \ and\
  \bibinfo {author} {\bibfnamefont {S.~A.}\ \bibnamefont {Teukolsky}},\ }\href
  {\doibase 10.1103/PhysRevD.82.124016} {\bibfield  {journal} {\bibinfo
  {journal} {Phys. Rev.}\ }\textbf {\bibinfo {volume} {D82}},\ \bibinfo {pages}
  {124016} (\bibinfo {year} {2010})},\ \Eprint {http://arxiv.org/abs/1004.4697}
  {arXiv:1004.4697 [gr-qc]} \BibitemShut {NoStop}%
\bibitem [{\citenamefont {Lousto}\ \emph {et~al.}(2016)\citenamefont {Lousto},
  \citenamefont {Healy},\ and\ \citenamefont {Nakano}}]{Lousto:2015uwa}%
  \BibitemOpen
  \bibfield  {author} {\bibinfo {author} {\bibfnamefont {C.~O.}\ \bibnamefont
  {Lousto}}, \bibinfo {author} {\bibfnamefont {J.}~\bibnamefont {Healy}}, \
  and\ \bibinfo {author} {\bibfnamefont {H.}~\bibnamefont {Nakano}},\ }\href
  {\doibase 10.1103/PhysRevD.93.044031} {\bibfield  {journal} {\bibinfo
  {journal} {Phys. Rev.}\ }\textbf {\bibinfo {volume} {D93}},\ \bibinfo {pages}
  {044031} (\bibinfo {year} {2016})},\ \Eprint
  {http://arxiv.org/abs/1506.04768} {arXiv:1506.04768 [gr-qc]} \BibitemShut
  {NoStop}%
\bibitem [{\citenamefont {Huerta}\ \emph {et~al.}(2017)\citenamefont {Huerta}
  \emph {et~al.}}]{Huerta:2016rwp}%
  \BibitemOpen
  \bibfield  {author} {\bibinfo {author} {\bibfnamefont {E.~A.}\ \bibnamefont
  {Huerta}} \emph {et~al.},\ }\href {\doibase 10.1103/PhysRevD.95.024038}
  {\bibfield  {journal} {\bibinfo  {journal} {Phys. Rev.}\ }\textbf {\bibinfo
  {volume} {D95}},\ \bibinfo {pages} {024038} (\bibinfo {year} {2017})},\
  \Eprint {http://arxiv.org/abs/1609.05933} {arXiv:1609.05933 [gr-qc]}
  \BibitemShut {NoStop}%
\bibitem [{\citenamefont {Kelly}\ \emph {et~al.}(2011)\citenamefont {Kelly},
  \citenamefont {Baker}, \citenamefont {Boggs}, \citenamefont {McWilliams},\
  and\ \citenamefont {Centrella}}]{Kelly:2011bp}%
  \BibitemOpen
  \bibfield  {author} {\bibinfo {author} {\bibfnamefont {B.~J.}\ \bibnamefont
  {Kelly}}, \bibinfo {author} {\bibfnamefont {J.~G.}\ \bibnamefont {Baker}},
  \bibinfo {author} {\bibfnamefont {W.~D.}\ \bibnamefont {Boggs}}, \bibinfo
  {author} {\bibfnamefont {S.~T.}\ \bibnamefont {McWilliams}}, \ and\ \bibinfo
  {author} {\bibfnamefont {J.}~\bibnamefont {Centrella}},\ }\href {\doibase
  10.1103/PhysRevD.84.084009} {\bibfield  {journal} {\bibinfo  {journal} {Phys.
  Rev.}\ }\textbf {\bibinfo {volume} {D84}},\ \bibinfo {pages} {084009}
  (\bibinfo {year} {2011})},\ \Eprint {http://arxiv.org/abs/1107.1181}
  {arXiv:1107.1181 [gr-qc]} \BibitemShut {NoStop}%
\bibitem [{\citenamefont {Hinderer}\ and\ \citenamefont
  {Babak}(2017)}]{Hinderer:2017jcs}%
  \BibitemOpen
  \bibfield  {author} {\bibinfo {author} {\bibfnamefont {T.}~\bibnamefont
  {Hinderer}}\ and\ \bibinfo {author} {\bibfnamefont {S.}~\bibnamefont
  {Babak}},\ }\href@noop {} {\  (\bibinfo {year} {2017})},\ \Eprint
  {http://arxiv.org/abs/1707.08426} {arXiv:1707.08426 [gr-qc]} \BibitemShut
  {NoStop}%
\bibitem [{\citenamefont {Cao}\ and\ \citenamefont {Han}(2017)}]{Cao:2017ndf}%
  \BibitemOpen
  \bibfield  {author} {\bibinfo {author} {\bibfnamefont {Z.}~\bibnamefont
  {Cao}}\ and\ \bibinfo {author} {\bibfnamefont {W.-B.}\ \bibnamefont {Han}},\
  }\href {\doibase 10.1103/PhysRevD.96.044028} {\bibfield  {journal} {\bibinfo
  {journal} {Phys. Rev.}\ }\textbf {\bibinfo {volume} {D96}},\ \bibinfo {pages}
  {044028} (\bibinfo {year} {2017})},\ \Eprint
  {http://arxiv.org/abs/1708.00166} {arXiv:1708.00166 [gr-qc]} \BibitemShut
  {NoStop}%
\bibitem [{sxs()}]{sxscatalogue}%
  \BibitemOpen
  \href@noop {} {}\bibinfo {howpublished}
  {\url{https://www.black-holes.org/waveforms/catalog.php}}\BibitemShut
  {NoStop}%
\bibitem [{ecc()}]{eccentricimrcode}%
  \BibitemOpen
  \href@noop {} {}\bibinfo {howpublished}
  {\url{https://github.com/ianhinder/EccentricIMR}}\BibitemShut {NoStop}%
\bibitem [{\citenamefont {Gopakumar}\ and\ \citenamefont
  {Iyer}(1997)}]{Gopakumar:1997bs}%
  \BibitemOpen
  \bibfield  {author} {\bibinfo {author} {\bibfnamefont {A.}~\bibnamefont
  {Gopakumar}}\ and\ \bibinfo {author} {\bibfnamefont {B.~R.}\ \bibnamefont
  {Iyer}},\ }\href {\doibase 10.1103/PhysRevD.56.7708} {\bibfield  {journal}
  {\bibinfo  {journal} {Phys. Rev.}\ }\textbf {\bibinfo {volume} {D56}},\
  \bibinfo {pages} {7708} (\bibinfo {year} {1997})},\ \Eprint
  {http://arxiv.org/abs/gr-qc/9710075} {arXiv:gr-qc/9710075} \BibitemShut
  {NoStop}%
\bibitem [{\citenamefont {Gopakumar}\ and\ \citenamefont
  {Iyer}(2002)}]{Gopakumar:2001dy}%
  \BibitemOpen
  \bibfield  {author} {\bibinfo {author} {\bibfnamefont {A.}~\bibnamefont
  {Gopakumar}}\ and\ \bibinfo {author} {\bibfnamefont {B.~R.}\ \bibnamefont
  {Iyer}},\ }\href {\doibase 10.1103/PhysRevD.65.084011} {\bibfield  {journal}
  {\bibinfo  {journal} {Phys. Rev.}\ }\textbf {\bibinfo {volume} {D65}},\
  \bibinfo {pages} {084011} (\bibinfo {year} {2002})},\ \Eprint
  {http://arxiv.org/abs/gr-qc/0110100} {arXiv:gr-qc/0110100} \BibitemShut
  {NoStop}%
\bibitem [{\citenamefont {Poisson}\ and\ \citenamefont
  {Will}(2014)}]{poisson2014gravity}%
  \BibitemOpen
  \bibfield  {author} {\bibinfo {author} {\bibfnamefont {E.}~\bibnamefont
  {Poisson}}\ and\ \bibinfo {author} {\bibfnamefont {C.}~\bibnamefont {Will}},\
  }\href
  {http://www.cambridge.org/academic/subjects/physics/cosmology-relativity-and-gravitation/gravity-newtonian-post-newtonian-relativistic}
  {\emph {\bibinfo {title} {Gravity: Newtonian, Post-Newtonian,
  Relativistic}}}\ (\bibinfo  {publisher} {Cambridge University Press},\
  \bibinfo {year} {2014})\BibitemShut {NoStop}%
\bibitem [{\citenamefont {Boyle}\ \emph {et~al.}(2007)\citenamefont {Boyle},
  \citenamefont {Brown}, \citenamefont {Kidder}, \citenamefont {Mrou\'{e}},
  \citenamefont {Pfeiffer}, \citenamefont {Scheel}, \citenamefont {Cook},\ and\
  \citenamefont {Teukolsky}}]{Boyle2007}%
  \BibitemOpen
  \bibfield  {author} {\bibinfo {author} {\bibfnamefont {M.}~\bibnamefont
  {Boyle}}, \bibinfo {author} {\bibfnamefont {D.}~\bibnamefont {Brown}},
  \bibinfo {author} {\bibfnamefont {L.}~\bibnamefont {Kidder}}, \bibinfo
  {author} {\bibfnamefont {A.}~\bibnamefont {Mrou\'{e}}}, \bibinfo {author}
  {\bibfnamefont {H.}~\bibnamefont {Pfeiffer}}, \bibinfo {author}
  {\bibfnamefont {M.}~\bibnamefont {Scheel}}, \bibinfo {author} {\bibfnamefont
  {G.}~\bibnamefont {Cook}}, \ and\ \bibinfo {author} {\bibfnamefont
  {S.}~\bibnamefont {Teukolsky}},\ }\href {\doibase 10.1103/PhysRevD.76.124038}
  {\bibfield  {journal} {\bibinfo  {journal} {Physical Review D}\ }\textbf
  {\bibinfo {volume} {76}},\ \bibinfo {pages} {1} (\bibinfo {year}
  {2007})}\BibitemShut {NoStop}%
\bibitem [{\citenamefont {MacDonald}\ \emph {et~al.}(2013)\citenamefont
  {MacDonald}, \citenamefont {Mroue}, \citenamefont {Pfeiffer}, \citenamefont
  {Boyle}, \citenamefont {Kidder}, \citenamefont {Scheel}, \citenamefont
  {Szilagyi},\ and\ \citenamefont {Taylor}}]{MacDonald:2012mp}%
  \BibitemOpen
  \bibfield  {author} {\bibinfo {author} {\bibfnamefont {I.}~\bibnamefont
  {MacDonald}}, \bibinfo {author} {\bibfnamefont {A.~H.}\ \bibnamefont
  {Mroue}}, \bibinfo {author} {\bibfnamefont {H.~P.}\ \bibnamefont {Pfeiffer}},
  \bibinfo {author} {\bibfnamefont {M.}~\bibnamefont {Boyle}}, \bibinfo
  {author} {\bibfnamefont {L.~E.}\ \bibnamefont {Kidder}}, \bibinfo {author}
  {\bibfnamefont {M.~A.}\ \bibnamefont {Scheel}}, \bibinfo {author}
  {\bibfnamefont {B.}~\bibnamefont {Szilagyi}}, \ and\ \bibinfo {author}
  {\bibfnamefont {N.~W.}\ \bibnamefont {Taylor}},\ }\href {\doibase
  10.1103/PhysRevD.87.024009} {\bibfield  {journal} {\bibinfo  {journal} {Phys.
  Rev.}\ }\textbf {\bibinfo {volume} {D87}},\ \bibinfo {pages} {024009}
  (\bibinfo {year} {2013})},\ \Eprint {http://arxiv.org/abs/1210.3007}
  {arXiv:1210.3007 [gr-qc]} \BibitemShut {NoStop}%
\bibitem [{SpE()}]{SpECwebsite}%
  \BibitemOpen
  \href@noop {} {}\bibinfo {howpublished}
  {\url{http://www.black-holes.org/SpEC.html}}\BibitemShut {NoStop}%
\bibitem [{\citenamefont {Scheel}\ \emph {et~al.}(2006)\citenamefont {Scheel},
  \citenamefont {Pfeiffer}, \citenamefont {Lindblom}, \citenamefont {Kidder},
  \citenamefont {Rinne},\ and\ \citenamefont {Teukolsky}}]{Scheel:2006gg}%
  \BibitemOpen
  \bibfield  {author} {\bibinfo {author} {\bibfnamefont {M.~A.}\ \bibnamefont
  {Scheel}}, \bibinfo {author} {\bibfnamefont {H.~P.}\ \bibnamefont
  {Pfeiffer}}, \bibinfo {author} {\bibfnamefont {L.}~\bibnamefont {Lindblom}},
  \bibinfo {author} {\bibfnamefont {L.~E.}\ \bibnamefont {Kidder}}, \bibinfo
  {author} {\bibfnamefont {O.}~\bibnamefont {Rinne}}, \ and\ \bibinfo {author}
  {\bibfnamefont {S.~A.}\ \bibnamefont {Teukolsky}},\ }\href {\doibase
  10.1103/PhysRevD.74.104006} {\bibfield  {journal} {\bibinfo  {journal} {Phys.
  Rev.}\ }\textbf {\bibinfo {volume} {D74}},\ \bibinfo {pages} {104006}
  (\bibinfo {year} {2006})},\ \Eprint {http://arxiv.org/abs/gr-qc/0607056}
  {arXiv:gr-qc/0607056 [gr-qc]} \BibitemShut {NoStop}%
\bibitem [{\citenamefont {Szilagyi}\ \emph {et~al.}(2009)\citenamefont
  {Szilagyi}, \citenamefont {Lindblom},\ and\ \citenamefont
  {Scheel}}]{Szilagyi:2009qz}%
  \BibitemOpen
  \bibfield  {author} {\bibinfo {author} {\bibfnamefont {B.}~\bibnamefont
  {Szilagyi}}, \bibinfo {author} {\bibfnamefont {L.}~\bibnamefont {Lindblom}},
  \ and\ \bibinfo {author} {\bibfnamefont {M.~A.}\ \bibnamefont {Scheel}},\
  }\href {\doibase 10.1103/PhysRevD.80.124010} {\bibfield  {journal} {\bibinfo
  {journal} {Phys. Rev.}\ }\textbf {\bibinfo {volume} {D80}},\ \bibinfo {pages}
  {124010} (\bibinfo {year} {2009})},\ \Eprint {http://arxiv.org/abs/0909.3557}
  {arXiv:0909.3557 [gr-qc]} \BibitemShut {NoStop}%
\bibitem [{\citenamefont {Buchman}\ \emph {et~al.}(2012)\citenamefont
  {Buchman}, \citenamefont {Pfeiffer}, \citenamefont {Scheel},\ and\
  \citenamefont {Szilagyi}}]{Buchman:2012dw}%
  \BibitemOpen
  \bibfield  {author} {\bibinfo {author} {\bibfnamefont {L.~T.}\ \bibnamefont
  {Buchman}}, \bibinfo {author} {\bibfnamefont {H.~P.}\ \bibnamefont
  {Pfeiffer}}, \bibinfo {author} {\bibfnamefont {M.~A.}\ \bibnamefont
  {Scheel}}, \ and\ \bibinfo {author} {\bibfnamefont {B.}~\bibnamefont
  {Szilagyi}},\ }\href {\doibase 10.1103/PhysRevD.86.084033} {\bibfield
  {journal} {\bibinfo  {journal} {Phys. Rev.}\ }\textbf {\bibinfo {volume}
  {D86}},\ \bibinfo {pages} {084033} (\bibinfo {year} {2012})},\ \Eprint
  {http://arxiv.org/abs/1206.3015} {arXiv:1206.3015 [gr-qc]} \BibitemShut
  {NoStop}%
\bibitem [{\citenamefont {Lindblom}\ \emph {et~al.}(2006)\citenamefont
  {Lindblom}, \citenamefont {Scheel}, \citenamefont {Kidder}, \citenamefont
  {Owen},\ and\ \citenamefont {Rinne}}]{Lindblom:2005qh}%
  \BibitemOpen
  \bibfield  {author} {\bibinfo {author} {\bibfnamefont {L.}~\bibnamefont
  {Lindblom}}, \bibinfo {author} {\bibfnamefont {M.~A.}\ \bibnamefont
  {Scheel}}, \bibinfo {author} {\bibfnamefont {L.~E.}\ \bibnamefont {Kidder}},
  \bibinfo {author} {\bibfnamefont {R.}~\bibnamefont {Owen}}, \ and\ \bibinfo
  {author} {\bibfnamefont {O.}~\bibnamefont {Rinne}},\ }\href {\doibase
  10.1088/0264-9381/23/16/S09} {\bibfield  {journal} {\bibinfo  {journal}
  {Class. Quant. Grav.}\ }\textbf {\bibinfo {volume} {23}},\ \bibinfo {pages}
  {S447} (\bibinfo {year} {2006})},\ \Eprint
  {http://arxiv.org/abs/gr-qc/0512093} {arXiv:gr-qc/0512093 [gr-qc]}
  \BibitemShut {NoStop}%
\bibitem [{\citenamefont {Friedrich}(1985)}]{Friedrich1985}%
  \BibitemOpen
  \bibfield  {author} {\bibinfo {author} {\bibfnamefont {H.}~\bibnamefont
  {Friedrich}},\ }\href {\doibase 10.1007/BF01217728} {\bibfield  {journal}
  {\bibinfo  {journal} {Communications in Mathematical Physics}\ }\textbf
  {\bibinfo {volume} {100}},\ \bibinfo {pages} {525} (\bibinfo {year}
  {1985})}\BibitemShut {NoStop}%
\bibitem [{\citenamefont {Pretorius}(2005)}]{Pretorius2005a}%
  \BibitemOpen
  \bibfield  {author} {\bibinfo {author} {\bibfnamefont {F.}~\bibnamefont
  {Pretorius}},\ }\href {\doibase 10.1103/PhysRevLett.95.121101} {\bibfield
  {journal} {\bibinfo  {journal} {Phys. Rev. Lett.}\ }\textbf {\bibinfo
  {volume} {95}},\ \bibinfo {pages} {121101} (\bibinfo {year} {2005})},\
  \Eprint {http://arxiv.org/abs/gr-qc/0507014} {arXiv:gr-qc/0507014 [gr-qc]}
  \BibitemShut {NoStop}%
\bibitem [{\citenamefont {Lindblom}\ and\ \citenamefont
  {Szilagyi}(2009)}]{Lindblom:2009tu}%
  \BibitemOpen
  \bibfield  {author} {\bibinfo {author} {\bibfnamefont {L.}~\bibnamefont
  {Lindblom}}\ and\ \bibinfo {author} {\bibfnamefont {B.}~\bibnamefont
  {Szilagyi}},\ }\href {\doibase 10.1103/PhysRevD.80.084019} {\bibfield
  {journal} {\bibinfo  {journal} {Phys. Rev.}\ }\textbf {\bibinfo {volume}
  {D80}},\ \bibinfo {pages} {084019} (\bibinfo {year} {2009})},\ \Eprint
  {http://arxiv.org/abs/0904.4873} {arXiv:0904.4873 [gr-qc]} \BibitemShut
  {NoStop}%
\bibitem [{\citenamefont {Hemberger}\ \emph {et~al.}(2013)\citenamefont
  {Hemberger}, \citenamefont {Scheel}, \citenamefont {Kidder}, \citenamefont
  {Szilágyi}, \citenamefont {Lovelace}, \citenamefont {Taylor},\ and\
  \citenamefont {Teukolsky}}]{Hemberger:2012jz}%
  \BibitemOpen
  \bibfield  {author} {\bibinfo {author} {\bibfnamefont {D.~A.}\ \bibnamefont
  {Hemberger}}, \bibinfo {author} {\bibfnamefont {M.~A.}\ \bibnamefont
  {Scheel}}, \bibinfo {author} {\bibfnamefont {L.~E.}\ \bibnamefont {Kidder}},
  \bibinfo {author} {\bibfnamefont {B.}~\bibnamefont {Szilágyi}}, \bibinfo
  {author} {\bibfnamefont {G.}~\bibnamefont {Lovelace}}, \bibinfo {author}
  {\bibfnamefont {N.~W.}\ \bibnamefont {Taylor}}, \ and\ \bibinfo {author}
  {\bibfnamefont {S.~A.}\ \bibnamefont {Teukolsky}},\ }\href {\doibase
  10.1088/0264-9381/30/11/115001} {\bibfield  {journal} {\bibinfo  {journal}
  {Class. Quant. Grav.}\ }\textbf {\bibinfo {volume} {30}},\ \bibinfo {pages}
  {115001} (\bibinfo {year} {2013})},\ \Eprint {http://arxiv.org/abs/1211.6079}
  {arXiv:1211.6079 [gr-qc]} \BibitemShut {NoStop}%
\bibitem [{\citenamefont {Scheel}\ \emph {et~al.}(2015)\citenamefont {Scheel},
  \citenamefont {Giesler}, \citenamefont {Hemberger}, \citenamefont {Lovelace},
  \citenamefont {Kuper}, \citenamefont {Boyle}, \citenamefont {Szilágyi},\
  and\ \citenamefont {Kidder}}]{Scheel:2014ina}%
  \BibitemOpen
  \bibfield  {author} {\bibinfo {author} {\bibfnamefont {M.~A.}\ \bibnamefont
  {Scheel}}, \bibinfo {author} {\bibfnamefont {M.}~\bibnamefont {Giesler}},
  \bibinfo {author} {\bibfnamefont {D.~A.}\ \bibnamefont {Hemberger}}, \bibinfo
  {author} {\bibfnamefont {G.}~\bibnamefont {Lovelace}}, \bibinfo {author}
  {\bibfnamefont {K.}~\bibnamefont {Kuper}}, \bibinfo {author} {\bibfnamefont
  {M.}~\bibnamefont {Boyle}}, \bibinfo {author} {\bibfnamefont
  {B.}~\bibnamefont {Szilágyi}}, \ and\ \bibinfo {author} {\bibfnamefont
  {L.~E.}\ \bibnamefont {Kidder}},\ }\href {\doibase
  10.1088/0264-9381/32/10/105009} {\bibfield  {journal} {\bibinfo  {journal}
  {Class. Quant. Grav.}\ }\textbf {\bibinfo {volume} {32}},\ \bibinfo {pages}
  {105009} (\bibinfo {year} {2015})},\ \Eprint {http://arxiv.org/abs/1412.1803}
  {arXiv:1412.1803 [gr-qc]} \BibitemShut {NoStop}%
\bibitem [{\citenamefont {Pfeiffer}\ \emph {et~al.}(2007)\citenamefont
  {Pfeiffer}, \citenamefont {Brown}, \citenamefont {Kidder}, \citenamefont
  {Lindblom}, \citenamefont {Lovelace},\ and\ \citenamefont
  {Scheel}}]{Pfeiffer:2007yz}%
  \BibitemOpen
  \bibfield  {author} {\bibinfo {author} {\bibfnamefont {H.~P.}\ \bibnamefont
  {Pfeiffer}}, \bibinfo {author} {\bibfnamefont {D.~A.}\ \bibnamefont {Brown}},
  \bibinfo {author} {\bibfnamefont {L.~E.}\ \bibnamefont {Kidder}}, \bibinfo
  {author} {\bibfnamefont {L.}~\bibnamefont {Lindblom}}, \bibinfo {author}
  {\bibfnamefont {G.}~\bibnamefont {Lovelace}}, \ and\ \bibinfo {author}
  {\bibfnamefont {M.~A.}\ \bibnamefont {Scheel}},\ }\bibfield  {booktitle}
  {\emph {\bibinfo {booktitle} {{New frontiers in numerical relativity.
  Proceedings, International Meeting, NFNR 2006, Potsdam, Germany, July 17-21,
  2006}}},\ }\href {\doibase 10.1088/0264-9381/24/12/S06} {\bibfield  {journal}
  {\bibinfo  {journal} {Class. Quant. Grav.}\ }\textbf {\bibinfo {volume}
  {24}},\ \bibinfo {pages} {S59} (\bibinfo {year} {2007})},\ \Eprint
  {http://arxiv.org/abs/gr-qc/0702106} {arXiv:gr-qc/0702106 [gr-qc]}
  \BibitemShut {NoStop}%
\bibitem [{\citenamefont {Szilágyi}(2014)}]{Szilagyi:2014fna}%
  \BibitemOpen
  \bibfield  {author} {\bibinfo {author} {\bibfnamefont {B.}~\bibnamefont
  {Szilágyi}},\ }\href {\doibase 10.1142/S0218271814300146} {\bibfield
  {journal} {\bibinfo  {journal} {Int. J. Mod. Phys.}\ }\textbf {\bibinfo
  {volume} {D23}},\ \bibinfo {pages} {1430014} (\bibinfo {year} {2014})},\
  \Eprint {http://arxiv.org/abs/1405.3693} {arXiv:1405.3693 [gr-qc]}
  \BibitemShut {NoStop}%
\bibitem [{\citenamefont {Boyle}\ and\ \citenamefont
  {Mroue}(2009)}]{Boyle:2009vi}%
  \BibitemOpen
  \bibfield  {author} {\bibinfo {author} {\bibfnamefont {M.}~\bibnamefont
  {Boyle}}\ and\ \bibinfo {author} {\bibfnamefont {A.~H.}\ \bibnamefont
  {Mroue}},\ }\href {\doibase 10.1103/PhysRevD.80.124045} {\bibfield  {journal}
  {\bibinfo  {journal} {Phys. Rev.}\ }\textbf {\bibinfo {volume} {D80}},\
  \bibinfo {pages} {124045} (\bibinfo {year} {2009})},\ \Eprint
  {http://arxiv.org/abs/0905.3177} {arXiv:0905.3177 [gr-qc]} \BibitemShut
  {NoStop}%
\bibitem [{\citenamefont {Hinder}\ \emph {et~al.}(2014)\citenamefont {Hinder}
  \emph {et~al.}}]{Hinder:2013oqa}%
  \BibitemOpen
  \bibfield  {author} {\bibinfo {author} {\bibfnamefont {I.}~\bibnamefont
  {Hinder}} \emph {et~al.},\ }\href {\doibase 10.1088/0264-9381/31/2/025012}
  {\bibfield  {journal} {\bibinfo  {journal} {Class. Quant. Grav.}\ }\textbf
  {\bibinfo {volume} {31}},\ \bibinfo {pages} {025012} (\bibinfo {year}
  {2014})},\ \Eprint {http://arxiv.org/abs/1307.5307} {arXiv:1307.5307 [gr-qc]}
  \BibitemShut {NoStop}%
\bibitem [{\citenamefont {Healy}\ \emph {et~al.}(2017)\citenamefont {Healy},
  \citenamefont {Lousto}, \citenamefont {Nakano},\ and\ \citenamefont
  {Zlochower}}]{Healy:2017zqj}%
  \BibitemOpen
  \bibfield  {author} {\bibinfo {author} {\bibfnamefont {J.}~\bibnamefont
  {Healy}}, \bibinfo {author} {\bibfnamefont {C.~O.}\ \bibnamefont {Lousto}},
  \bibinfo {author} {\bibfnamefont {H.}~\bibnamefont {Nakano}}, \ and\ \bibinfo
  {author} {\bibfnamefont {Y.}~\bibnamefont {Zlochower}},\ }\href {\doibase
  10.1088/1361-6382/aa7929} {\bibfield  {journal} {\bibinfo  {journal} {Class.
  Quant. Grav.}\ }\textbf {\bibinfo {volume} {34}},\ \bibinfo {pages} {145011}
  (\bibinfo {year} {2017})},\ \Eprint {http://arxiv.org/abs/1702.00872}
  {arXiv:1702.00872 [gr-qc]} \BibitemShut {NoStop}%
\bibitem [{\citenamefont {Bohé}\ \emph {et~al.}(2017)\citenamefont {Bohé}
  \emph {et~al.}}]{Bohe:2016gbl}%
  \BibitemOpen
  \bibfield  {author} {\bibinfo {author} {\bibfnamefont {A.}~\bibnamefont
  {Bohé}} \emph {et~al.},\ }\href {\doibase 10.1103/PhysRevD.95.044028}
  {\bibfield  {journal} {\bibinfo  {journal} {Phys. Rev.}\ }\textbf {\bibinfo
  {volume} {D95}},\ \bibinfo {pages} {044028} (\bibinfo {year} {2017})},\
  \Eprint {http://arxiv.org/abs/1611.03703} {arXiv:1611.03703 [gr-qc]}
  \BibitemShut {NoStop}%
\bibitem [{\citenamefont {Blackman}\ \emph {et~al.}(2015)\citenamefont
  {Blackman}, \citenamefont {Field}, \citenamefont {Galley}, \citenamefont
  {Szilágyi}, \citenamefont {Scheel}, \citenamefont {Tiglio},\ and\
  \citenamefont {Hemberger}}]{Blackman:2015pia}%
  \BibitemOpen
  \bibfield  {author} {\bibinfo {author} {\bibfnamefont {J.}~\bibnamefont
  {Blackman}}, \bibinfo {author} {\bibfnamefont {S.~E.}\ \bibnamefont {Field}},
  \bibinfo {author} {\bibfnamefont {C.~R.}\ \bibnamefont {Galley}}, \bibinfo
  {author} {\bibfnamefont {B.}~\bibnamefont {Szilágyi}}, \bibinfo {author}
  {\bibfnamefont {M.~A.}\ \bibnamefont {Scheel}}, \bibinfo {author}
  {\bibfnamefont {M.}~\bibnamefont {Tiglio}}, \ and\ \bibinfo {author}
  {\bibfnamefont {D.~A.}\ \bibnamefont {Hemberger}},\ }\href {\doibase
  10.1103/PhysRevLett.115.121102} {\bibfield  {journal} {\bibinfo  {journal}
  {Phys. Rev. Lett.}\ }\textbf {\bibinfo {volume} {115}},\ \bibinfo {pages}
  {121102} (\bibinfo {year} {2015})},\ \Eprint
  {http://arxiv.org/abs/1502.07758} {arXiv:1502.07758 [gr-qc]} \BibitemShut
  {NoStop}%
\bibitem [{\citenamefont {Finn}(1992)}]{Finn:1992wt}%
  \BibitemOpen
  \bibfield  {author} {\bibinfo {author} {\bibfnamefont {L.~S.}\ \bibnamefont
  {Finn}},\ }\href {\doibase 10.1103/PhysRevD.46.5236} {\bibfield  {journal}
  {\bibinfo  {journal} {Phys. Rev.}\ }\textbf {\bibinfo {volume} {D46}},\
  \bibinfo {pages} {5236} (\bibinfo {year} {1992})},\ \Eprint
  {http://arxiv.org/abs/gr-qc/9209010} {arXiv:gr-qc/9209010 [gr-qc]}
  \BibitemShut {NoStop}%
\bibitem [{O1P()}]{O1PSD}%
  \BibitemOpen
  \href@noop {} {}\bibinfo {howpublished}
  {\url{https://github.com/ligo-cbc/pycbc-config/blob/master/O1/psd/H1L1-ER8_HARM_MEAN_PSD-1126033217-223200.txt.gz}}\BibitemShut
  {NoStop}%
\bibitem [{\citenamefont {Usman}\ \emph {et~al.}(2016)\citenamefont {Usman}
  \emph {et~al.}}]{Usman:2015kfa}%
  \BibitemOpen
  \bibfield  {author} {\bibinfo {author} {\bibfnamefont {S.~A.}\ \bibnamefont
  {Usman}} \emph {et~al.},\ }\href {\doibase 10.1088/0264-9381/33/21/215004}
  {\bibfield  {journal} {\bibinfo  {journal} {Class. Quant. Grav.}\ }\textbf
  {\bibinfo {volume} {33}},\ \bibinfo {pages} {215004} (\bibinfo {year}
  {2016})},\ \Eprint {http://arxiv.org/abs/1508.02357} {arXiv:1508.02357
  [gr-qc]} \BibitemShut {NoStop}%
\bibitem [{\citenamefont {Shoemaker}(2010)}]{Shoemaker:2010}%
  \BibitemOpen
  \bibfield  {author} {\bibinfo {author} {\bibfnamefont {D.}~\bibnamefont
  {Shoemaker}} (\bibinfo {collaboration} {{LIGO} Collaboration}),\ }\href
  {https://dcc.ligo.org/LIGO-T0900288-v3/public} {\enquote {\bibinfo {title}
  {Advanced {LIGO} anticipated sensitivity curves},}\ } (\bibinfo {year}
  {2010}),\ \bibinfo {note} {{LIGO} Document T0900288-v3}\BibitemShut {NoStop}%
\bibitem [{\citenamefont {Hansen}(2014)}]{hansen:fourier}%
  \BibitemOpen
  \bibfield  {author} {\bibinfo {author} {\bibfnamefont {E.~W.}\ \bibnamefont
  {Hansen}},\ }\href@noop {} {\emph {\bibinfo {title} {Fourier Transforms:
  Principles and Applications}}}\ (\bibinfo  {publisher} {Wiley},\ \bibinfo
  {year} {2014})\BibitemShut {NoStop}%
\bibitem [{\citenamefont {McKechan}\ \emph {et~al.}(2010)\citenamefont
  {McKechan}, \citenamefont {Robinson},\ and\ \citenamefont
  {Sathyaprakash}}]{McKechan:2010kp}%
  \BibitemOpen
  \bibfield  {author} {\bibinfo {author} {\bibfnamefont {D.~J.~A.}\
  \bibnamefont {McKechan}}, \bibinfo {author} {\bibfnamefont {C.}~\bibnamefont
  {Robinson}}, \ and\ \bibinfo {author} {\bibfnamefont {B.~S.}\ \bibnamefont
  {Sathyaprakash}},\ }\bibfield  {booktitle} {\emph {\bibinfo {booktitle}
  {{Gravitational waves. Proceedings, 8th Edoardo Amaldi Conference, Amaldi 8,
  New York, USA, June 22-26, 2009}}},\ }\href {\doibase
  10.1088/0264-9381/27/8/084020} {\bibfield  {journal} {\bibinfo  {journal}
  {Class. Quant. Grav.}\ }\textbf {\bibinfo {volume} {27}},\ \bibinfo {pages}
  {084020} (\bibinfo {year} {2010})},\ \Eprint {http://arxiv.org/abs/1003.2939}
  {arXiv:1003.2939 [gr-qc]} \BibitemShut {NoStop}%
\bibitem [{\citenamefont {Taracchini}(2017)}]{taracchini:fpeak}%
  \BibitemOpen
  \bibfield  {author} {\bibinfo {author} {\bibfnamefont {A.}~\bibnamefont
  {Taracchini}},\ }\href@noop {} {}\bibinfo {howpublished} {private
  communication} (\bibinfo {year} {2017})\BibitemShut {NoStop}%
\bibitem [{\citenamefont {Husa}\ \emph {et~al.}(2016)\citenamefont {Husa},
  \citenamefont {Khan}, \citenamefont {Hannam}, \citenamefont {Pürrer},
  \citenamefont {Ohme}, \citenamefont {Jiménez~Forteza},\ and\ \citenamefont
  {Bohé}}]{Husa:2015iqa}%
  \BibitemOpen
  \bibfield  {author} {\bibinfo {author} {\bibfnamefont {S.}~\bibnamefont
  {Husa}}, \bibinfo {author} {\bibfnamefont {S.}~\bibnamefont {Khan}}, \bibinfo
  {author} {\bibfnamefont {M.}~\bibnamefont {Hannam}}, \bibinfo {author}
  {\bibfnamefont {M.}~\bibnamefont {Pürrer}}, \bibinfo {author} {\bibfnamefont
  {F.}~\bibnamefont {Ohme}}, \bibinfo {author} {\bibfnamefont {X.}~\bibnamefont
  {Jiménez~Forteza}}, \ and\ \bibinfo {author} {\bibfnamefont
  {A.}~\bibnamefont {Bohé}},\ }\href {\doibase 10.1103/PhysRevD.93.044006}
  {\bibfield  {journal} {\bibinfo  {journal} {Phys. Rev.}\ }\textbf {\bibinfo
  {volume} {D93}},\ \bibinfo {pages} {044006} (\bibinfo {year} {2016})},\
  \Eprint {http://arxiv.org/abs/1508.07250} {arXiv:1508.07250 [gr-qc]}
  \BibitemShut {NoStop}%
\bibitem [{\citenamefont {Hinder}\ and\ \citenamefont
  {Wardell}()}]{SimulationToolsWeb}%
  \BibitemOpen
  \bibfield  {author} {\bibinfo {author} {\bibfnamefont {I.}~\bibnamefont
  {Hinder}}\ and\ \bibinfo {author} {\bibfnamefont {B.}~\bibnamefont
  {Wardell}},\ }\href {http://simulationtools.org} {\enquote {\bibinfo {title}
  {{SimulationTools} for {Mathematica}},}\ }\bibinfo {note}
  {\url{http://simulationtools.org}}\BibitemShut {NoStop}%
\end{thebibliography}%

\end{document}